\documentstyle[psfig]{mn}

\newif\ifAMStwofonts
\newcommand{\q}{${\bf q}$}
\newcommand{\x}{${\bf x}$}
\newcommand{\s}{${\bf S}$}
\newcommand{\pot}{$\varphi$}
\newcommand{\mq}{{\bf q}}
\newcommand{\mx}{{\bf x}}
\newcommand{\ms}{{\bf S}}
\newcommand{\be}{\begin{equation}}
\newcommand{\ee}{\end{equation}}
\newcommand{\bea}{\begin{eqnarray}}
\newcommand{\eea}{\end{eqnarray}}
\ifoldfss
  \ifCUPmtlplainloaded \else
    \NewTextAlphabet{textbfit} {cmbxti10} {}
    \NewTextAlphabet{textbfss} {cmssbx10} {}
    \NewMathAlphabet{mathbfit} {cmbxti10} {} % for math mode
    \NewMathAlphabet{mathbfss} {cmssbx10} {} %  "   "    "
  \fi
  \ifAMStwofonts
    \ifCUPmtlplainloaded \else
      \NewSymbolFont{upmath} {eurm10}
      \NewSymbolFont{AMSa} {msam10}
      \NewMathSymbol{\upi}     {0}{upmath}{19}
      \NewMathSymbol{\umu}     {0}{upmath}{16}
      \NewMathSymbol{\upartial}{0}{upmath}{40}
      \NewMathSymbol{\leqslant}{3}{AMSa}{36}
      \NewMathSymbol{\geqslant}{3}{AMSa}{3E}

      \let\leq=\leqslant 
      \let\geq=\geqslant 
    \fi
  \fi
\fi % End of OFSS

\ifnfssone
  \newmathalphabet{\mathit}
  \addtoversion{normal}{\mathit}{cmr}{m}{it}
  \addtoversion{bold}{\mathit}{cmr}{bx}{it}
  \newmathalphabet{\mathbfit} % math mode version of \textbfit{..}
  \addtoversion{normal}{\mathbfit}{cmr}{bx}{it}
  \addtoversion{bold}{\mathbfit}{cmr}{bx}{it}
  \newmathalphabet{\mathbfss} % math mode version of \textbfss{..}
  \addtoversion{normal}{\mathbfss}{cmss}{bx}{n}
  \addtoversion{bold}{\mathbfss}{cmss}{bx}{n}
  \ifAMStwofonts
    \ifCUPmtlplainloaded \else
      \UseAMStwoboldmath
      \makeatletter
      \new@mathgroup\upmath@group
      \define@mathgroup\mv@normal\upmath@group{eur}{m}{n}
      \define@mathgroup\mv@bold\upmath@group{eur}{b}{n}
      \edef\UPM{\hexnumber\upmath@group}
      \new@mathgroup\amsa@group
      \define@mathgroup\mv@normal\amsa@group{msa}{m}{n}
      \define@mathgroup\mv@bold\amsa@group{msa}{m}{n}
      \edef\AMSa{\hexnumber\amsa@group}
      \makeatother
      \mathchardef\upi="0\UPM19
      \mathchardef\umu="0\UPM16
      \mathchardef\upartial="0\UPM40
      \mathchardef\leqslant="3\AMSa36
      \mathchardef\geqslant="3\AMSa3E

      \let\leq=\leqslant 
      \let\geq=\geqslant 
    \fi
  \fi
\fi % End of NFSS release 1

\ifnfsstwo
  \DeclareMathAlphabet{\mathbfit}{OT1}{cmr}{bx}{it}
  \SetMathAlphabet\mathbfit{bold}{OT1}{cmr}{bx}{it}
  \DeclareMathAlphabet{\mathbfss}{OT1}{cmss}{bx}{n}
  \SetMathAlphabet\mathbfss{bold}{OT1}{cmss}{bx}{n}
  \ifAMStwofonts
    \ifCUPmtlplainloaded \else
      \DeclareSymbolFont{UPM}{U}{eur}{m}{n}
      \SetSymbolFont{UPM}{bold}{U}{eur}{b}{n}
      \DeclareSymbolFont{AMSa}{U}{msa}{m}{n}
      \DeclareMathSymbol{\upi}{0}{UPM}{"19}
      \DeclareMathSymbol{\umu}{0}{UPM}{"16}
      \DeclareMathSymbol{\upartial}{0}{UPM}{"40}
      \DeclareMathSymbol{\leqslant}{3}{AMSa}{"36}
      \DeclareMathSymbol{\geqslant}{3}{AMSa}{"3E}

      \let\leq=\leqslant 
      \let\geq=\geqslant 
    \fi
  \fi
\fi % End of NFSS release 2

\ifCUPmtlplainloaded \else
  \ifAMStwofonts \else % If no AMS fonts
    \def\upi{\pi}
    \def\umu{\mu}
    \def\upartial{\partial}
  \fi
\fi

\title[Mass Function Theory: I Dynamics]{A Lagrangian Dynamical Theory 
       for the Mass Function\\ of Cosmic Structures: I Dynamics}

\author[P. Monaco]
       {Pierluigi Monaco\\
        Scuola Internazionale Superiore di Studi Avanzati (SISSA), via  
        Beirut 4, 34014 -- Trieste, Italy \\
        Dipartimento di Astronomia, Universit\`a degli studi di Trieste, via 
        Tiepolo 11, 34131 -- Trieste, Italy\\
	Email: monaco@sissa.it}
 
\date{}

\pagerange{\pageref{firstpage}--\pageref{lastpage}}
\pubyear{1996}

\begin{document}

\maketitle

\label{firstpage}

\begin{abstract}
A new theory for determining the mass function of cosmic structures is
presented. It relies on a realistic treatment of collapse dynamics.
Gravitational collapse is analyzed in the Lagrangian perturbative
framework. Lagrangian perturbations provide an approximation of
truncated type, i.e. small-scale structure is filtered out.  The
collapse time is suitably defined as the instant at which orbit
crossing takes place. The convergence of the Lagrangian series in
predicting the collapse time of a homogeneous ellipsoid is
demonstrated; it is also shown that third-order calculations are
necessary in predicting collapse. Then, the Lagrangian prediction,
with a correction for quasi-spherical perturbations, can be used to
determine the collapse time of a homogeneous ellipsoid in a fast and
precise way. Furthermore, ellipsoidal collapse can be considered as a
particular truncation of the Lagrangian series. Gaussian fields with
scale-free power spectra are then considered. The Lagrangian series
for the collapse time is found to converge when the collapse time is
not large.  In this case, ellipsoidal collapse gives a fast and
accurate approximation of the collapse time; spherical collapse is
found to poorly reproduce the collapse time, even in a statistical
sense. Analytical fits of the distribution functions of the inverse
collapse times, as predicted by the ellipsoid model and by third-order
Lagrangian theory, are given. These will be necessary for a
determination of the mass function, which will be given in paper II.
\end{abstract}


%%%%%%%%%%%%%%%%%%%%%%%%%%%%%  1  %%%%%%%%%%%%%%%%%%%%%%%%%%%%%%%%%%

\section{Introduction}

An important outcome of any cosmological model is the distribution of
the masses of those collapsed clumps which form during gravitational
collapse; this quantity is usually called mass function (hereafter
MF), or multiplicity function. These structures correspond to the
observed galaxies, groups and clusters of galaxies.  Thanks to recent
efforts, the masses of these observed cosmic structures can be
estimated, though with large uncertainties and relying on uncertain
hypothesis. Galaxy cluster masses can be estimated by means of three
different methods: estimates based on optical galaxies, such as virial
estimates (Biviano et al. 1993; see also Bahcall \& Cen 1993), X-ray
temperatures (Henry \& Arnaud 1991), and
gravitational lensing (see, e.g., Fort \& Mellier 1994). Every method
has its drawbacks: virial estimates rely on the delicate hypothesis of
virial equilibrium; X-ray analyses rely on safer gas-dynamical
hypotheses but are confined to the inner regions of the cluster;
lensing estimates directly probe the gravitational potential, but are
now available only for a few clusters.  Group masses are much more
difficult to estimate, due to the small number of galaxies involved
and to the uncertain dynamical status of groups. Virial mass
estimates, corrected for unvirialization, are available for a large
number of groups (Pisani et al. 1992).  Galaxy masses are relatively
easier to determine, due to the more evolved dynamical status and to
the large number of tracers (stars instead of galaxies) (see Ashman,
Persic \& Salucci 1993). However, mass estimates are mainly limited to
a few optical radii, while dark-matter halos probably extend further.

The state of the art of the MF theory presents severe problems as
well.  Cosmological structures are the sites of highly non-linear
dynamics; it is well known that, for generic initial conditions, the
Newtonian collapse of a general self-gravitating system has no known
solution in the highly non-linear regime. Analytical approximations
and N-body simulations can help to face the problem in an approximate
way. In this paper I will mainly focus on realistic analytical
approximations to gravitational collapse, needed to develop a MF
theory. Another paper (Monaco 1996, hereafter paper II) develops the
statistical tools necessary to get a MF.

Despite the intrinsic difficulties, a heuristic solution of the MF
problem was found as early as 1974 by Press \& Schechter (1974;
hereafter PS). I have already described in full detail the PS MF in a
previous paper (Monaco 1995; hereafter M95). It suffices now to recall
that all the dynamical difficulties of non-linear dynamics are
circumvented by assuming that `something happens' to a mass clump
(i.e. a fully virialized structure forms) as soon as its linearly
evolved density contrast $\delta$ reaches a given threshold $\delta_c$
of order one. One of the simplest fully non-linear collapse models,
the spherical model, can be invoked to give a more precise
determination of the quantity $\delta_c$: indeed, the spherical model
predicts full collapse to a singularity as soon as the linearly
extrapolated density contrast reaches $\delta_c$=1.69.  Surprisingly,
the PS formula has been found to reproduce in a more or less accurate
way the results of large N-body simulations (see references and
discussions in M95).  Comparisons of the PS MF to N-body simulations
have often suggested a lower value for the $\delta_c$ parameter, even
though technical details, such as the group-finding algorithm and the
shape of the filter function, have to be taken into account to obtain
a precise number. The original PS formulation has been improved in a
number of papers; in particular, Peacock \& Heavens (1990) and Bond et
al. (1991) have solved the so-called `cloud-in-cloud' problem, while
Bower (1991) and Lacey \& Cole (1993, 1994) have extended the
formalism to describing merging histories of dark matter halos. A more
complete list of references and a deep discussion on these points is
given in paper II.

It is worth stressing that the only dynamical ingredients of the PS
recipe are linear theory and a fixed density threshold. Thus, the PS
MF can only be considered as a fitting formula which nicely describes
the results of many N-body simulations. A great advantage in using
only linear theory is the following: as linear theory predicts the
density field to be rescaled, as it evolves, by a factor which depends
only on time, the statistical properties of the final field are the
same as the initial field, which is usually assumed to be a Gaussian
process.  In practice, evolved density fields develop strong and
non-trivial non-Gaussian features.  Nonetheless, most works on the MF,
since that of PS, have detailed the statistical treatment of regions
with initial density larger than a given threshold (called `excursion
sets'), or peaks of the initial density field, rarely detailing the
dynamical description of collapse.

Only a few authors have considered more evolved dynamics than linear
theory.  Lucchin \& Matarrese (1988) and Porciani et al. (1996), in
two different ways, introduced non-Gaussianity into the MF. In Lucchin
\& Matarrese (1988) the non-Gaussianity was explicitly introduced in a
PS-like approach; it could be either primordial or of dynamical
origin. Porciani et al. (1996) introduced non-Gaussianity in the
framework of diffusion theory (Bond et al. 1991), by putting a
reflecting barrier at $\delta=-1$, to avoid unphysical negative
densities. In both cases, gravitationally-induced non-Gaussianity was
found to cause enhanced production of high-mass clumps, and, in
Porciani et al. (1996), an intriguing cut-off at low masses was
predicted. Vergassola et al. (1994) calculated the MF in the framework
of adhesion theory; collapsed structures were identified with
caustics.  Another attempt to model in a more realistic way the
formation of collapsed clumps was made by Bond \& Myers (1996), in the
framework of peak-patch theory. In that theory, the collapse times of
the peak-patches were estimated by means of the homogeneous
ellipsoidal collapse model, and the final locations were found by
means of Zel'dovich (1970) approximation. Ellipsoidal collapse in a
cosmological context was used also by Eisenstein \& Loeb (1995) to
estimate the angular momentum of collapsing structures.

A very different approach was used by Cavaliere and collaborators in a
number of works (see Cavaliere, Menci \& Tozzi 1994 for a recent
review). First they constructed a MF theory based on dynamical
timescales, then they used a kinetic approach to model the aggregation
of already collapsed clumps, and finally they found a formalism, based
on Cayley trees, which was able to unify the kinetic and diffusion
approaches.  They recently applied this formalism to adhesion theory
(Cavaliere, Menci \& Tozzi 1996). Similar or related approaches were
recently used by Shaviv \& Shaviv (1995) and by Sheth (1996).  Such
approaches, very different from the ones mentioned above, attempt to
describe the highly non-linear behaviour of matter clumps after first
collapse.

In a previous paper, M95, I attempted to introduce realistic dynamics
in the framework of a PS-like theory. The PS idea of identifying the
fraction of collapsed matter as the probability that a point fulfills
certain collapse conditions was used to determine the MF.  It was
shown that such kind of approach greatly simplifies if `local'
predictions of collapse are used, i.e. if the collapse of a mass
element is based on some relevant quantities {\it relative only to the
element considered}. Collapse predictions were estimated by means of
ansatze based on the Zel'dovich approximation and by means of the
homogeneous ellipsoid collapse model. The result was an increase in
large-mass objects with respect to the usual PS prediction, which
could explain why lower $\delta_c$ values than the spherical 1.69 are
found when comparing the PS MF to N-body simulations (lower values of
$\delta_c$ correspond to a shift of the large-mass tail of the MF
toward large masses).

The M95 work can be seen as a simple `variation on a theme' of the PS
procedure, plagued by all the PS faults, e.g. the cloud-in-cloud
problem.  Anyway, that variation suffices in shedding light on a
number of problems which have a precise dynamical meaning, and which
are connected to the use of the spherical collapse model.  To be more
precise, the spherical model predicts full collapse of the whole
structure to a singularity at a given time; it is usually assumed that
at that time the structure fully virializes. In this way: (i) collapse
is quite well defined and (ii) virialization surely occurs (iii) at the
same moment as the collapse, so that the two concepts can be used
equivalently.  In practice, when collapse takes place with a realistic
geometry, (i) it has to be carefully defined, and different authors
have in fact different ideas on what collapse is (e.g., is the `real'
collapse of a homogeneous ellipsoid that on the first axis, or that on
all three axes?); (ii) while it is somehow possible to model collapse,
it is terribly difficult to model virialization, and it is terribly
difficult to decide when and where virialization is going to have
place (clusters and groups of galaxies are generally {\it not}
virialized!).

In this paper, and in paper II, the ideas contained in M95 are pushed
further, in order to construct a complete theory of the mass
distributions of cosmic structures, based on realistic dynamics.  This
paper focuses on the dynamical estimate of the collapse time of a mass
element, and on the definition of collapse. Paper II will be concerned
with the statistics necessary to get a MF from the definition of
collapse and its distribution. The whole set-up of the dynamical
problem is based on the idea, already formulated in M95, that the MF
is an intrinsically Lagrangian quantity, in the sense that it is best
faced within the framework of Lagrangian fluidodynamics. In Section 2
the machinery of Lagrangian dynamics of a cosmological
self-gravitating cold fluid is presented. Lagrangian perturbation
theory is introduced: it provides an excellent framework to estimate
the collapse time of a mass element. In Section 3 the definition of
collapse is analyzed, its punctual, non-local nature is stressed, and
a way of defining collapse in the case of non-filtered fields is
presented; this can be implemented as the collapse definition when
comparing this theory to N-body simulations. In Section 4 the
Lagrangian perturbative series, up to the 3rd order, is successfully
applied to the homogeneous ellipsoid collapse.  In Section 5 the
collapse of a Gaussian field with power-law spectra, generated in
cubic grids of 32$^3$ points, is carefully analyzed. As a result, the
simple ellipsoidal model is found to give a very accurate estimate of
the collapse time, at least for the largest objects which
collapse. The distribution of the inverses of collapse times is
analyzed and quantified; this quantity is necessary for the
calculation of the MF. Section 6 contains a summary and conclusions.
Technical details about Lagrangian perturbations and ellipsoidal
collapse are given in two Appendices.

%%%%%%%%%%%%%%%%%%%%%%%%%%%%%  2  %%%%%%%%%%%%%%%%%%%%%%%%%%%%%%%%%%

\section{The Lagrangian Nature of the Mass Function} 

The MF is the distribution of the masses of the collapsed, isolated
clumps of mass. To get a MF, information about the mass elements
which undergo collapse is needed; it is not necessary to know where
the collapsed clumps go to. Then, the MF problem is suitably analyzed
in a Lagrangian fluidodynamical framework, where the independent space
variable \q\ is related to a given mass element, whose motion is
followed. This is different from the usual Eulerian picture, where the
space variable \x\ is related to a given spatial point, and the
evolution of the elements, which instantaneously happen to be in that
point, is followed.  This is the starting point of the MF theory
presented here. Recently, the same point of view has been taken by
Audit \& Alimi (1996).

Consider a pressureless, unvortical, Newtonian, self-gravitating
fluid in a perturbed Friedmann-Robertson-Walker Universe, with given
cosmological parameters $\Omega_0$ and $\Lambda$. The trajectory \x\
of every (infinitely small) mass element, initially at the comoving
position \q, can be written as:

\be \mx(\mq,t) = \mq + \ms(\mq,t)\; . \label{eq:mapping}\ee

\noindent \s\ is the displacement field (in comoving coordinates).
All the kinematic quantities relative to the mass element and its
density contrast can be expressed in terms of the displacement \s\
(see Appendix A, equation \ref{eq:kine}).  Note that, in this
approach, the density is not a dynamical quantity, as it was in the
Eulerian approach; the only dynamical quantity here is the
displacement field \s . It is possible to find evolution equations for
\s , equivalent to the usual Euler-Poisson system for the Newtonian
evolution of perturbations. Several authors, namely Buchert (1989),
Bouchet et al. (1995), Lachi\`eze-Rey (1993b) and Catelan (1995), have
written different equivalent forms of the same system of equations for
\s. The equations of Catelan (1995) are reported in Appendix A. In the
following it will always be assumed that the initial velocity field is
irrotational and parallel to the gravitational acceleration; see
Appendix A for further details.

It is possible to give initial conditions for the system through the
initial peculiar rescaled gravitational potential $\varphi(\mq,t_0)$,
instead of the usual initial density contrast $\delta(\mq,t_0)$; the
two fields are simply related by a Poisson equation:

\be \nabla^2 \varphi(\mq,t_0) = \frac{\delta(\mq,t_0)}{b(t_0)}\; ,
\label{eq:poi} \ee

\noindent
where $b(t)$ is the linear growing mode (note that $b(t_0) \simeq
a(t_0)$).

An important quantity is the Jacobian determinant of the transformation
given in equation (\ref{eq:mapping}):

\be J(\mq,t)=\det\left(\frac{\partial x_a}{\partial q_b}\right) = 
\det(\delta_{ab} + S_{a,b})  \label{eq:detjac}\ee

\noindent (comma denotes \q-derivative); the quantity $S_{a,b}$ is
commonly called {\it deformation tensor}.  The Lagrangian approach has
a natural limit in the condition $J=0$.  Before this moment, the
Lagrangian-to-Eulerian mapping is single-valued (single-stream or
laminar regime), i.e. mass elements coming from different points of
(Lagrangian) space do not arrive at the same Eulerian position. When
$J=0$ (orbit crossing or shell crossing, hereafter OC) the $\mq
\rightarrow \mx$ mapping becomes multi-valued (multi-stream regime),
and the density, being proportional to the inverse of $J$, goes to
infinity (see Shandarin \& Zel'dovich 1989 for a discussion of caustic
formation). It is beyond the scope of this paper to discuss the
possibility of using a Lagrangian approach in the multi-stream regime;
see Buchert (1994) for a discussion. Here it has to be noted that OC
takes place as soon as the first objects collapse, which, for any
realistic power spectrum, happens very soon after recombination. So
the Lagrangian formulation is, as it stands, essentially useless
unless the initial (potential or density) field is smoothed to
truncate the small-scale part of the power spectrum. This is a key
point: from a field $\varphi(\mq)$, which can in principle have power
on all scales, a hierarchy of smooth fields is worked out:

\be \varphi(\mq) \rightarrow \varphi(\mq;R_f) \; ,\ee

\noindent where $R_f$ is the width of the filter (the two fields are
both freely called \pot , but they are in fact different mathematical
objects). This highlights the strongest hypothesis of this theory,
namely that small-scale structure does not influence in a significant
way the dynamical evolution of larger scales before OC.  All the
following considerations apply to a general element of the
$\varphi(\mq;R_f)$ hierarchy, i.e. to a smoothed version of \pot; the
parameter $R_f$ will be omitted for simplicity's sake. (Note that some
approximation schemes, such as adhesion theory (Gurbatov, Saichev \&
Shandarin 1989) or frozen flow approximation (Matarrese et al. 1992),
avoid OC; in this case no smoothing of the initial field is needed in
principle.)

The Lagrangian evolution equations for \s , as well as the Eulerian
evolution equations for $\delta$, have not been solved in general
cases. Nonetheless, it is possible to find, as in the Eulerian case,
perturbative solutions. In the Eulerian case (see, e.g., Bouchet
1996), the equations are expanded in terms proportional to powers of
the density contrast; if this quantity is small, then successive
perturbative terms are increasingly smaller and the series converges.
As a consequence, the validity of the Eulerian perturbative scheme is
limited to small density contrasts. In the Lagrangian case, the
perturbed quantity is not the density, which is not a dynamical
quantity, but the (comoving) displacements of the particles from their
initial positions.  It is easy to understand why this simple change of
perturbing parameter causes dramatic improvements in the performances
of the approximations: the density, being proportional to the inverse
of $J$ (equation \ref{eq:kine}), becomes infinite when $J$ goes to
0. In this case, at least some matrix element of the deformation
tensor (e.g. one eigenvalue, if $S_{a,b}$ is symmetric) is of order
(minus) one, i.e. just at the limit of validity of the perturbation
scheme. In other words, when Lagrangian perturbations start to break
down, the density can have reached very large values.

A number of authors have analyzed the Lagrangian perturbation scheme
at various orders, up to the third (Buchert 1989; Moutarde et
al. 1991; Bouchet et al 1992; Buchert 1992; Buchert \& Ehlers 1993;
Lachi\`eze-Rey 1993a,b; Bouchet et al. 1995; Buchert 1994; Catelan
1995; Bouchet 1996; Buchert 1996). A brief list of the main results is
given in Appendix A; see the references for further details. As a
matter of fact, different authors use very different notations; the
notation I will use is similar to that of Catelan (1995).

The perturbative series for \s\ can be written, up to third order, as:

\bea \lefteqn{\ms(\mq,t) = b_1(t) \ms^{(1)}(\mq) +
b_2(t)\ms^{(2)}(\mq)+b_{3a}(t) \ms^{(3a)}(\mq)}
\nonumber\\ && + b_{3b}(t)\ms^{(3b)}(\mq) + b_{3c}(t)
\ms^{(3c)}(\mq) + \ldots\; . \eea

\noindent In the following, among models with $\Lambda\neq 0$, only
the flat ones will be considered (the others are not of great
cosmological interest). The time functions $b_n(t)$ are given in
Appendix A, equations (\ref{eq:tpert}). The first-order time function
is just (minus) the linear growing mode, $b_1(t)=-b(t)$, while the
others are, at leading order and with great accuracy (exactly for an
Einstein-de Sitter background), proportional to $b^2$ or $b^3$,
according to their order.  This means that, in all the calculations
that follow, the dependence on the background cosmology can be
factorized out by using $b(t)$ as time variable.  The spatial
equations for the $\ms^{(n)}(\mq)$ terms are Poisson equations; this
reflects the implicit non-locality of Newtonian gravitational
dynamics. Again, they are reported in Appendix A. Note that the
$\ms^{(2)}$, $\ms^{(3a)}$ and $\ms^{(3b)}$ terms are irrotational
($S^{(n)}_{a,b}$ is symmetric), while the $\ms^{(3c)}$ term is purely
rotational($S^{(3c)}_{a,b}$ is antisymmetric).

It is easy to recognize that the linear term of the perturbation,

\be \mx = \mq - b(t)\nabla\varphi\; , \ee

\noindent is the well-known Zel'dovich (1970) approximation. This is
not the place to list all the features, merits and limits of this
approximation; the reviews of Shandarin \& Zel'dovich (1989) and Sahni
\& Coles (1995) give full details and complete reference lists. Some
comments were reported also in M95. Is is worth mentioning that
Zel'dovich and 2nd order truncated approximations have been found very
successful in predicting the evolution of a N-body matter field in the
weakly non-linear regime, according to cross-correlation tests (Coles,
Melott \& Shandarin 1993; Melott, Buchert \& Wei\ss\ 1995; Sahni \&
Coles 1995). The third order has not been found to increase
significantly the precision of the series; moreover, third-order terms
are very sensitive to numerical errors, so the use of third-order
predictions is not generally recommended in that context.  Another
general conclusion is that the 3c term does not have great influence
on the density evolution (see also Buchert et al. 1997), which is to
be expected, as this term, being purely rotational, corresponds to a
rotation of the mass element in Lagrangian space, which does not
influence the density by itself.

As already noted in M95, there is a key difference between the usual
applications of the truncated Lagrangian approximations and the one
which is to be used here. In the works cited above the density fields
are evolved up to mass variances of order 1 or slightly more; at this
level the convergence of the Lagrangian series is more or less
guaranteed by construction. But the MF needs a prediction of collapse,
when all the perturbative terms become of the same order. The
convergence of the series in this case is not guaranteed and has to be
checked. This will be done in Section 4 in the case of ellipsoidal
collapse, and in Section 5 in the case of Gaussian fields with
scale-free power spectra.

%%%%%%%%%%%%%%%%%%%%%%%%%%%%%%%  3  %%%%%%%%%%%%%%%%%%%%%%%%%%%%%%%%%%%%%

\section{The definition of collapse}

The key quantity for the determination of a MF is the instant at which
a given mass element collapses. This raises the not trivial question
of what collapse means. Within the Lagrangian perturbation
framework, there is a natural definition of collapse:

\be J(\mq,b_c)=0 \label{eq:col} \ee

\noindent (note that the linear growing mode $b$ is the time variable;
$b_c$ is the collapse time). This instant has already been defined
before as OC.  At this instant a number of things happen: (i) the
density becomes infinite; (ii) the other kinematic quantities become
infinite (see M95); (iii) different trajectories intersect; (iv)
multi-stream regions form; (v) shock waves can form in a (subdominant)
dissipative component (baryons); (vi) gravitational dynamics becomes
interesting and (vii) really difficult to follow.  After OC a number
of other things can happen: violent relaxation, virialization, gas
cooling, star formation, supernovae feedback etc. All these relevant
events are decisive in making real astrophysical structures, but are
very difficult to model. Surely OC is a necessary condition for these
events to take place.

The main interest of the present theory is to model dark matter
clumps, not astrophysical objects. Dark matter clumps can be thought
of as local high-density concentrations of matter, regardless of their
geometry, internal dynamical status (relaxation, virialization or
otherwise) and so on.  Then the above definition of collapse is
meaningful, provided a mass element which has undergone collapse
remains in some high density clump; I assume that this is the
case. This assumption is reasonable: were it not the case, mass
elements entering a structure would evaporate back into the background
soon after; such evaporation events are assumed to be rare.

It is opportune not to introduce conditions on the internal dynamical
status of a collapsed clump at the level of collapse definition, for
at least two good reasons: (i) dynamics in the multi-stream regime is
not well understood, and any oversimplification, such as virialization
soon after collapse, can be misleading. (ii) Real objects, as groups
and clusters of galaxies, have plausibly undergone some kind of
(violent) relaxation, but their state is very complicated; only the
cores of rich groups and clusters are observed in a more evolved
state. Furthermore, a collapsed but not dynamically evolved clump may
contain smaller, more relaxed or fully virialized clumps, which can
survive in the larger structure for a significant time; this could be
the situation of galaxies in groups.  This kind of MF theory cannot
provide a description of such subclumps, or, in other words, cannot
represent galaxies in groups or clusters. A kinetic approach like that
described in Cavaliere et al. (1994) can model this kind of objects.

Another key feature of this definition is that it is clearly {\it
punctual}, in the sense that any prediction is relative to a point. It
is often assumed in the literature that the collapse prediction of a
point is to be extended to a surrounding region of filter-width size,
as the prediction is based on a mean over a given region; this fact,
reasonable especially when top-hat smoothing is used, leads to some
complications in the MF, as Blanchard, Valls-Gabaud \& Mamon (1992)
and Yano, Nagashima \& Gouda (1996) have shown. This is not my point
of view: filtering of the initial field is only necessary to suppress
small-scale orbit-crossed regions. All the kinematic and dynamical
quantities, and then the collapse predictions, are strictly punctual,
defined on vanishingly small mass elements. On the other hand, if the
field is smooth on a scale $R_f$, the collapse predictions will show a
coherence on a scale $R_f$, without any further assumptions.

Summing up, the definition of collapse given by equation
(\ref{eq:col}) can reasonably reproduce high-density clumps,
regardless of their internal dynamical status. Due to the truncated
nature of the approximations used, collapsed regions are predicted to
reach infinite density, but actual collapsing regions contain
smaller-scale structures which spread the collapsing mass around, thus
lowering the true density.  It is worthwhile wondering how a clump,
which is predicted to collapse when the field is smoothed at a given
scale, can be recognized in the unsmoothed evolved system, e.g., in a
N-body simulation.  Any conventional clump-finding algorithm, based on
percolation or overdensities above a given threshold, will probably
find more easily those collapsed regions which have certain
geometries, e.g.  spherical rather than filamentary.  The choice of
the clump-finding algorithm is decisive in this case, and it is
opportune to choose an algorithm based on the collapse definition
given above, equation (\ref{eq:col}). A good choice is to construct an
algorithm based simply on the concept of mapping, as infinite density
is only a consequence. I propose the following formula as an
implementation of equation (\ref{eq:col}):

\bea \lefteqn{\exists \mq_1, \mq_2\, :\ \ |\mq_1-\mq_2|\geq L,}\label{eq:map} 
\\ && |\mx(\mq_1,b)-\mx(\mq_2,b)|<\varepsilon,\ \ \ \ \varepsilon \ll L \; ,
\nonumber \eea

\noindent where both $\mq_1$ and $\mq_2$ are collapsed, relative to a
scale $L$, at the time $b>b_c$. This definition is very easy to
implement in an N-body simulation.

Obviously, a tight correspondence between high-density clumps and
regions found with equation (\ref{eq:map}) is expected, especially for
the most massive clumps. Nonetheless, the precise relationship has to
be checked with N-body simulations. In any case, the `objects'
predicted by this kind of theory are interesting in themselves, as
they are the sites of highly non-linear dynamics, and surely contain
the mass which is going to form astrophysical and cosmological
objects.  If more restricted classes of objects are needed, the
possibility of getting them by placing further constraints on the
collapsing point would be worth exploring.  As a conclusion, to be
conservative the MF which can be found with such a definition of
collapse could be called the MF of {\it high non-linearity
environments}.

%%%%%%%%%%%%%%%%%%%%%%%%%%%%%%%%  4  %%%%%%%%%%%%%%%%%%%%%%%%%%%%%%%%%%%%%%%

\section{Lagrangian perturbations in homogeneous ellipsoidal collapse}

As a first step, the gravitational collapse of a homogeneous triaxial
ellipsoid is considered. This exercise is useful for testing the
convergence of the Lagrangian series at OC, which is not guarenteed by
construction. Consider the potential:

\be \varphi(\mq) = \frac{1}{2}(\lambda_1 q_1^2 + \lambda_2 q_2^2 +
\lambda_3 q_3^2) \; ,\label{eq:ell} \ee

\noindent where the $\lambda_i$ are the eigenvalues of the first-order
deformation tensor $S^{(1)}_{a,b} = \varphi_{,ab}$; note the
difference of sign from the definition in M95. They are ordered as
follows: $\lambda_1 \geq \lambda_2 \geq \lambda_3$.  Because of
Poisson equation (\ref{eq:poi}), $\delta_l \equiv \delta(t_0)/b_0 =
\lambda_1 + \lambda_2 + \lambda_3 = {\rm const}$.  Equation (\ref{eq:ell})
represents the potential of a homogeneous ellipsoid in its principal
reference frame.

With this potential it is possible to solve all the Poisson equations
for the perturbing potentials. All the details of the calculations are
contained Appendix B. The solutions can be directly found by using
the so-called `local forms' of the perturbing potentials, which can be
found in Buchert \& Ehlers (1993), Buchert (1994) and Catelan
(1995). Briefly, it is possible to find vector fields, functions of
the initial potential and its derivatives {\it in the point
considered}, which solve the spatial equations (\ref{eq:spert}), but
are generally not irrotational. The local form $\ms^{(2L)}$ of the
second-order displacement is:

\be \ms^{(2)} = [{\bf \nabla}\varphi(\nabla^2\varphi)
- ({\bf \nabla}\varphi\cdot{\bf \nabla}){\bf \nabla}\varphi] 
+ {\bf R}^{(2)} = \ms^{(2L)} + {\bf R}^{(2)}\; . \label{eq:loc} \ee

\noindent The vector ${\bf R}^{(2)}$, divergenceless, is added to the
local form $\ms^{(2L)}$ in order to keep the vector $\ms^{(2)}$
irrotational; it contains the deep non-locality of gravitational
dynamics.  Analogous expressions for the 3a, 3b and 3c contributions
are reported in Appendix B.  The local forms are solutions of equations
(\ref{eq:spert}) if they are irrotational by themselves; this happens
only for a restricted class of initial conditions, to which the
homogeneous ellipsoidal potential belongs. Then the local forms can be
used to promptly obtain the perturbing potentials of the ellipsoid.

The 2nd-order local contribution to the deformation tensor,
$S^{(2L)}_{a,b}$, contains terms with second derivatives of \pot\ and
mixed terms with first and third derivatives. In the homogeneous
ellipsoidal case only the second derivative terms survive.  The same
happens for the 3aL and 3bL terms, while the 3cL term is null (see
Appendix B).  I call {\it ellipsoidal parts} those terms of the local
contributions to the deformation tensor which survive in the
ellipsoidal case:

\bea 
S^{(2E)}_{a,b} & = & \varphi_{,ab}\varphi_{,cc} - \varphi_{,ac}
\varphi_{,bc} \nonumber\\
S^{(3aE)}_{a,b} & = & \varphi_{,ac}\varphi_{,bc}^C \label{eq:elp} \\
S^{(3bE)}_{a,b} & = & \frac{1}{2}[S^{(2E)}_{ab}\varphi_{,cc} - S^{(2E)}_{bc}
\varphi_{,ac} + \varphi_{,ab}S^{(2E)}_{c,c} - \varphi_{,bc}S^{(2E)}_{a,c}]
\nonumber\\ 
S^{(3cE)}_{a,b} & = & 0\; ; \nonumber
\eea 

\noindent $\varphi_{,ab}^C$ is the cofactor matrix of $\varphi_{,ab}$.
These terms can be considered as a truncation of the local form, when
all derivatives of \pot\ greater than the second are neglected.  (Note
that 2nd-order local and ellipsoidal displacements, $\ms^{2L}$ and
$\ms^{2E}$, are equal, as the local form contains up to second
derivatives of the initial potential; the differences appear in the
deformation tensor. Third-order local and ellipsoidal displacements
are instead different.)

If the very weak $\Omega$ dependence of the time functions is
neglected (see equations \ref{eq:tpert}), the collapse equation $J=0$
is simply an algebraic equation in the variable $b(t)$ (the growing
mode), of order equal to the Lagrangian order.  All the technical
details are reported in Appendix B.  In the spherical case, in which
$\lambda_1 = \lambda_2 = \lambda_3 = \delta_l/3$, the collapse times
predicted by the Lagrangian series up to the 1st-, 2nd- and 3rd-order,
are $b_c\delta_l$ = 3, 2.27, 2.05. The exact $\Omega=1$ solution is
$b_c=a_c=1.69$; it can be appreciated that the results converge to the
exact value: the difference is reduced from 1.31 to .36. Other
cosmologies lead to very similar values of $b_c$, which differ from by
not more than 3 per cent (Lilje 1992).  For the Lagrangian
perturbation scheme, spherical symmetry is the hardest to deal with,
so we expect faster convergence in more general cases (see the
discussion in M95).

The first-order solution for general ellipsoids is simply $b_c^{(1)}
=1/\lambda_1$; it has been amply discussed in M95 (note that, due to
the different signs of the $\lambda_i$ eigenvalues, $b_c$ was
$-1/\lambda_3$ in M95).  The second-order solution for initially
overdense ellipsoids is:

\be b_c^{(2)} = \frac{7\lambda_1 - \sqrt{7 \lambda_1(\lambda_1+6\delta_l)}}
{3\lambda_1(\lambda_1-\delta_l)}\; . \label{eq:el2} \ee

\noindent If $\delta_l<0$, the second-order equation gives meaningful
solutions only if $\delta_l \geq -\lambda_1/6$, i.e.  only for
relatively small underdensities.  Other solutions exist which make
even a spherical void collapse! These had already been noted by Sahni
\& Shandarin (1996), and show how the 2nd-order Lagrangian scheme is
unreliable for initially underdense perturbations. The reason for this
behaviour is that 2nd-order perturbations, being of even order, do not
properly recognize voids, making them collapse.

The third-order solution is the smallest non-negative solution of the
equation:

\bea \lefteqn{1-\lambda_1 b_c^{(3)} - \frac{3}{14} \lambda_1 
(\delta_l-\lambda_1)(b_c^{(3)})^2} \\ && - \left( \frac{\mu_3}{126} + 
\frac{5}{84}\lambda_1\delta_l(\delta_l-\lambda_1)\right) (b_c^{(3)})^3 = 0\; ,
\nonumber\eea

\noindent where $\mu_3 = \lambda_1\lambda_2\lambda_3$. Though it is
not straightforward to choose analytically the right root of this
equation, it is very easy to find it with a computer. The third-order
equation gives meaningful solutions also for initially underdense
elements, thus avoiding the problems encountered with the second-order
solutions. This shows that third-order perturbation theory is
necessary when predicting the instant of collapse of general mass
elements.

Both at second and third-order, the first axis to collapse is the one
corresponding to the largest $\lambda$ eigenvalue, $\lambda_1$; this
is by itself an indication of convergence, as it means that the
first-order (Zel'dovich) approximation, which predicts collapse first
on the 1-axis, makes the greatest contribution to collapse dynamics.

The calculations of the exact collapse time of the ellipsoid have been
performed as in M95; the equations are reported in Appendix B.  To
improve the calculations, the integration has been divided into two
parts: after decoupling, defined as the instant when the density
starts to grow, the integration variable has been changed from
(logarithm of) $b$ to (logarithm of) the density; the integration has
been stopped at $\delta=\exp(15)$. This had led to a modest but
appreciable improvement in precision. The precision of the calculation
depends on the initial conditions: the spherical collapse is predicted
with a precision of around 0.2 per cent, while the `pancake' collapse,
when $\lambda_2=\lambda_3=0$, is recovered with a precision of 8 per
cent. The calculations have been performed only for $\delta_l$=1 or
$-1$; the other cases can be found by using the scaling relation:

\be b_c(\lambda_1,\lambda_2,\lambda_3) = kb_c(k\lambda_1,
k\lambda_2,k\lambda_3)\; . \ee

\begin{figure*}
(a) \hfill (b)
\centerline{
\psfig{figure=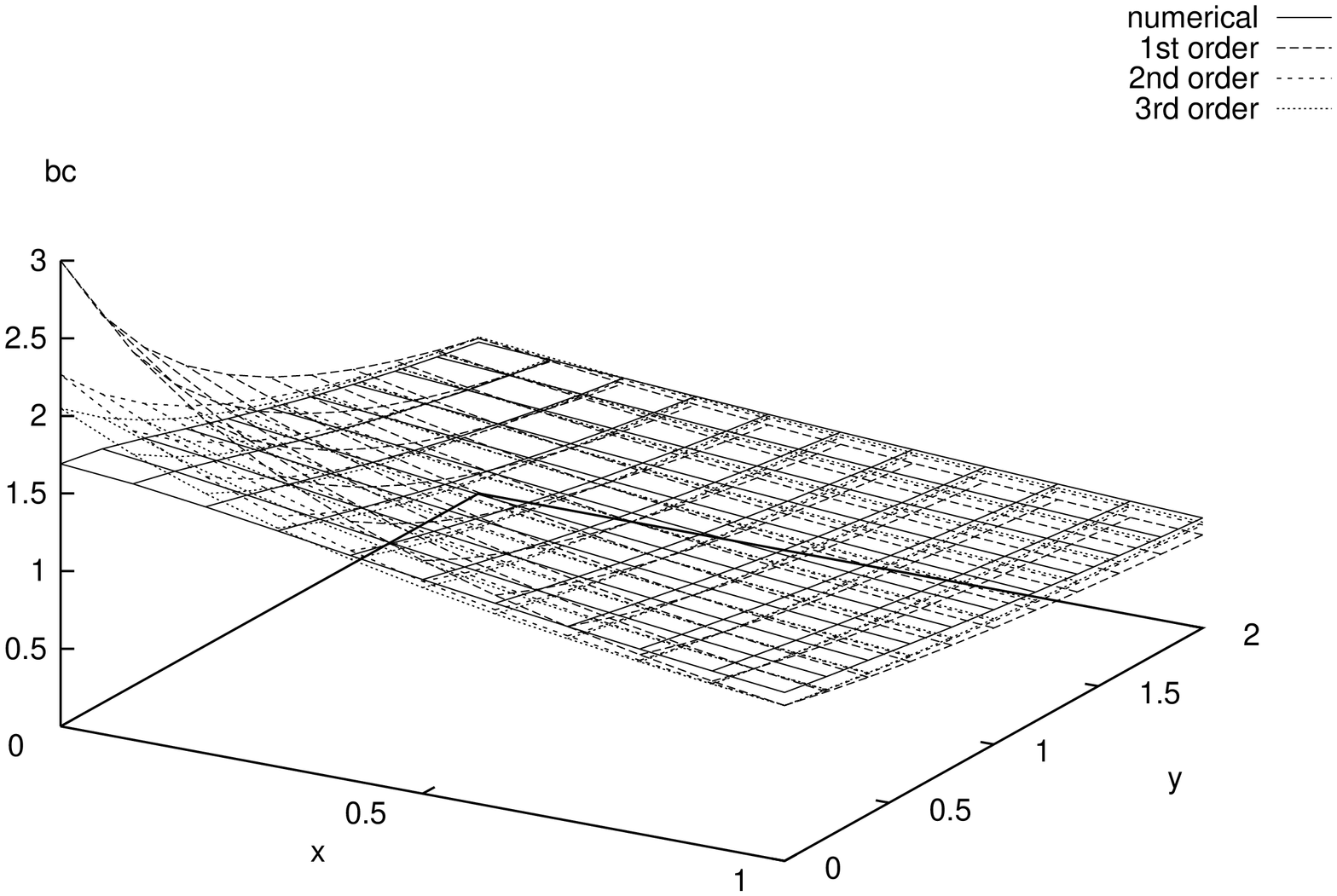,width=9.5cm,angle=-90}
\psfig{figure=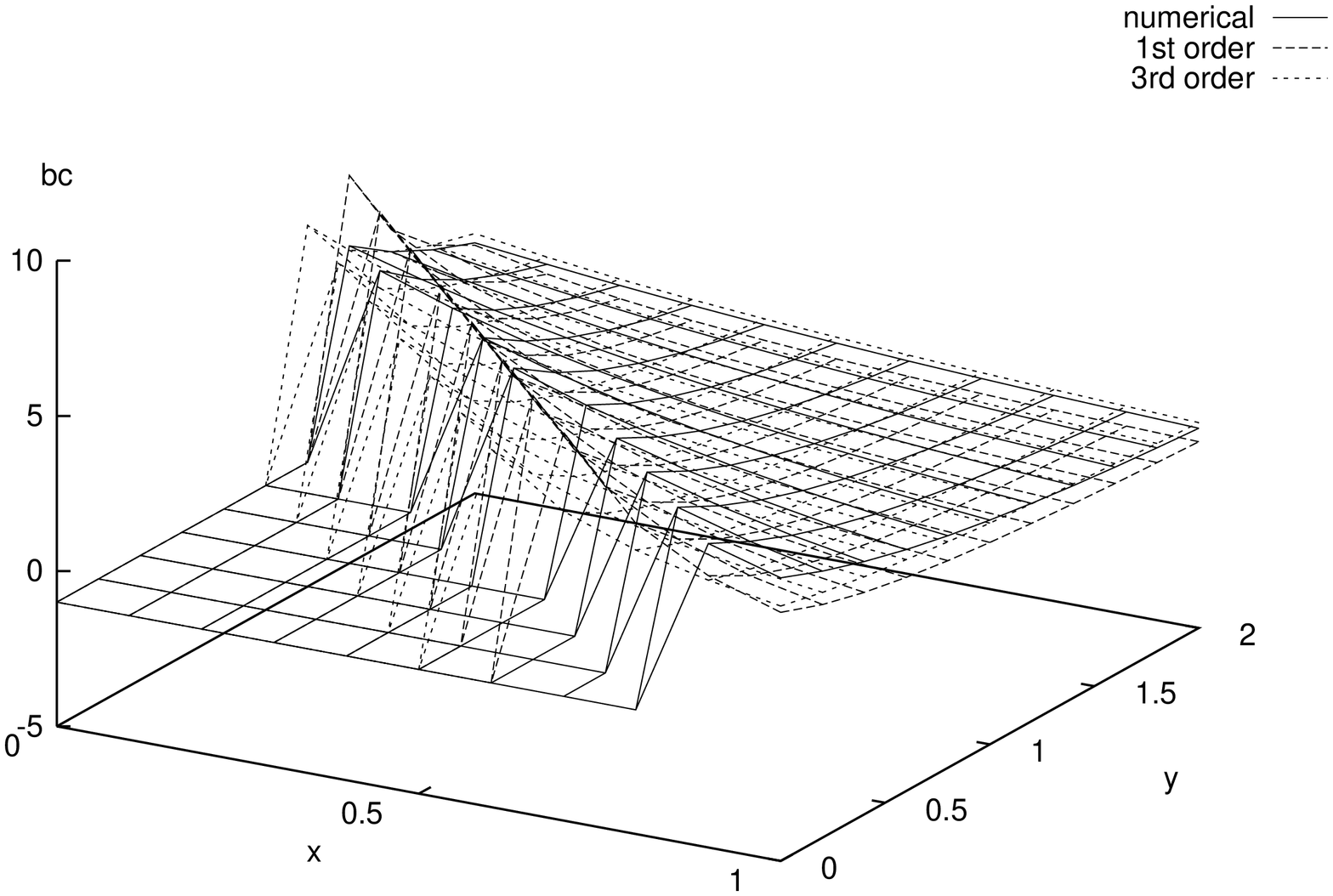,width=9.5cm,angle=-90}
}
(c) \hfill (d)
\centerline{
\psfig{figure=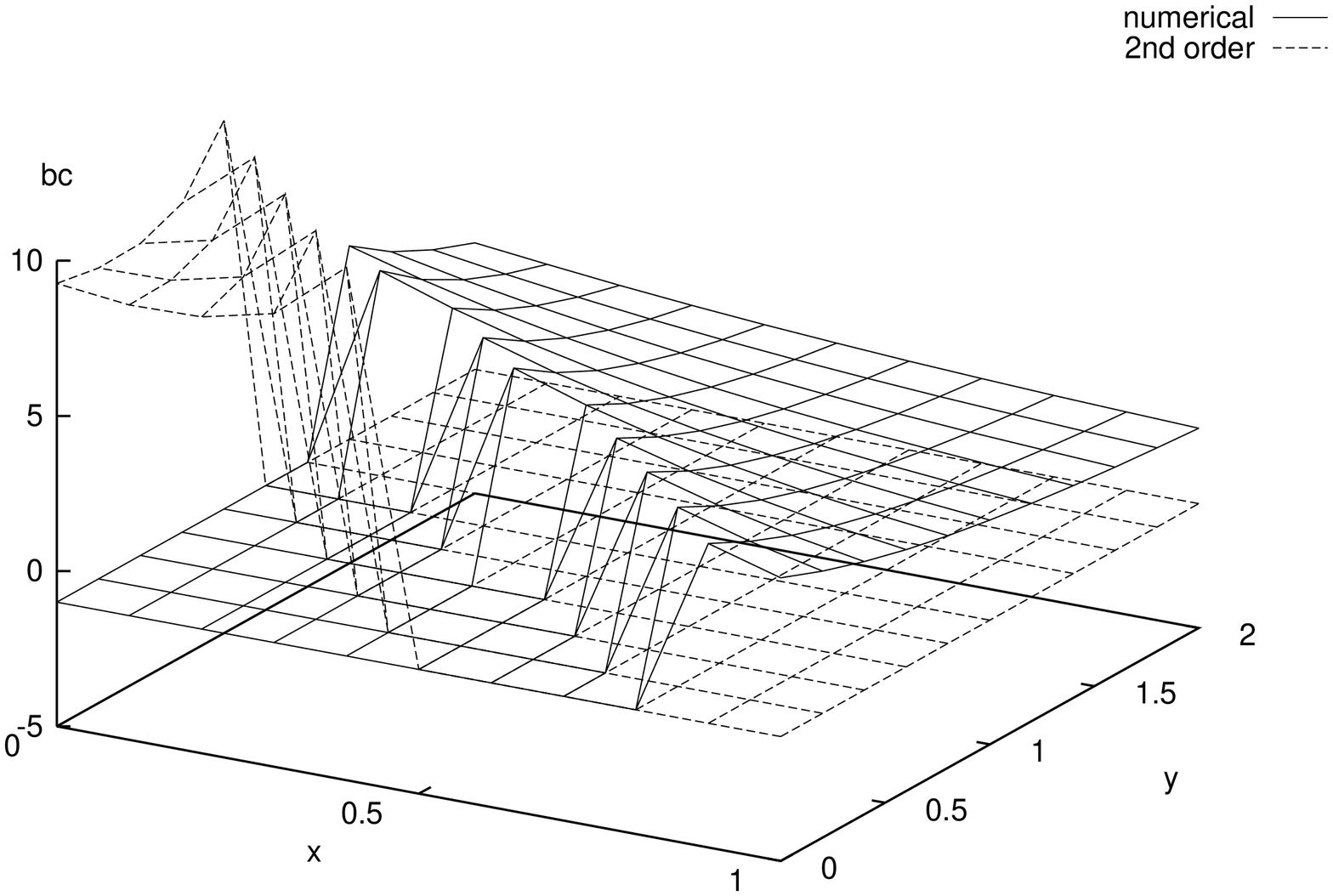,width=9.5cm,angle=-90}
\psfig{figure=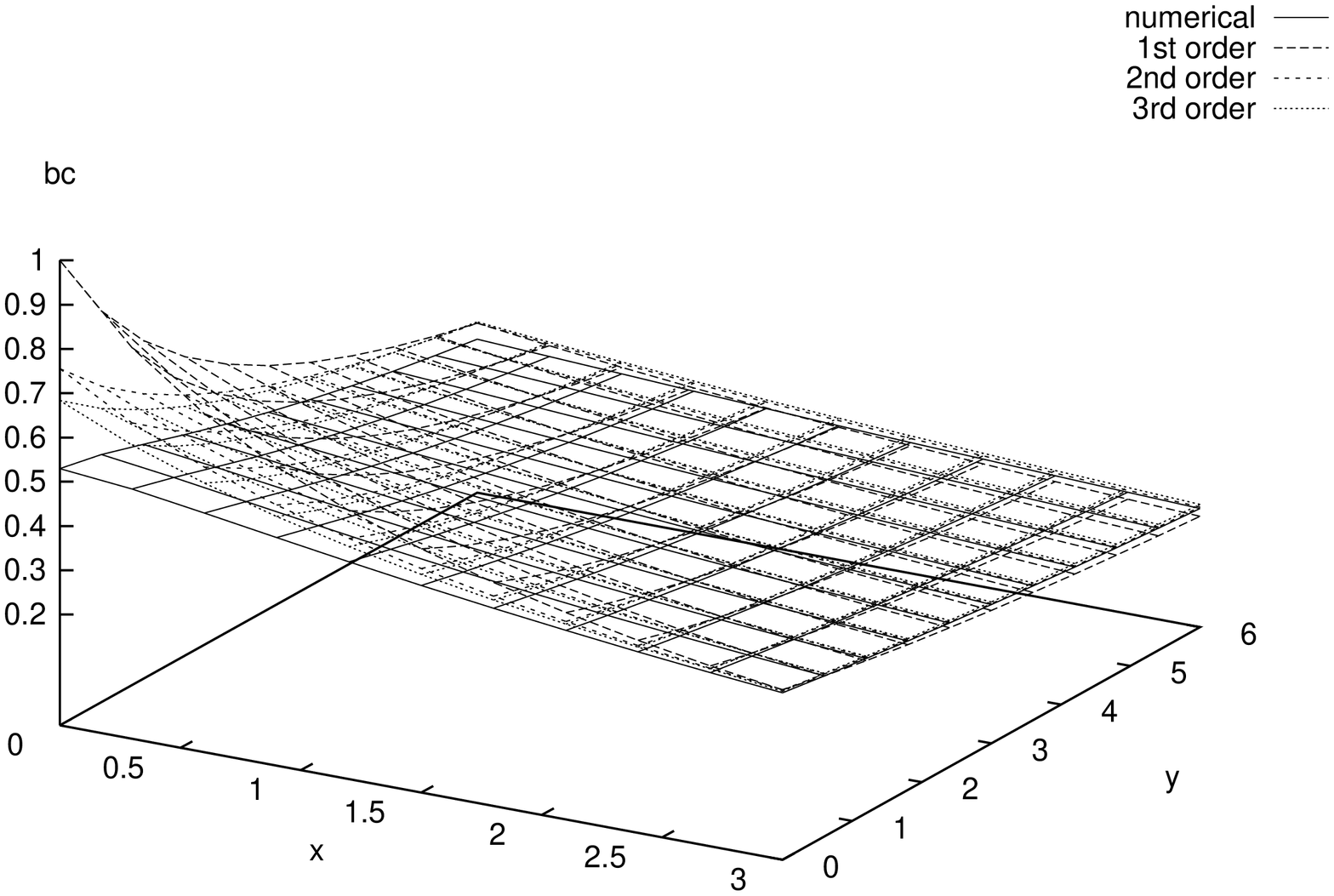,width=9.5cm,angle=-90}
}
\caption{Collapse times $b_c$ of homogeneous ellipsoids
according to Lagrangian perturbation theory, compared with a numerical
integration; $x=\lambda_1-\lambda_2$ and $y=\lambda_2-\lambda_3$. (a):
initial overdensity $\delta_l=1$.(b) and (c): initial underdensity
$\delta_l=-1$; 2nd-order predictions have been put in a separate
figure for the sake of clarity. (d): open universe, initial overdensity
$\delta_l=3$; the $\Omega$ dependence of the time functions is
neglected.}
\end{figure*}

Figs. 1a, b and c show the various $b_c$ curves of initially overdense
and underdense ellipsoids. The $x$ and $y$ variables, which range from
0 to $\infty$, are defined as $x=\lambda_1-\lambda_2$, $y=\lambda_2-
\lambda_3$; they give a measure of the initial shear of the ellipsoid
(see M95). A number of things can be noted:

\begin{enumerate}
\item The predictions at increasing Lagrangian orders always
converge to the exact value (within the numerical errors quoted
above); in the initial underdensity case, only the odd Lagrangian
orders converge to the solutions.
\item The convergence is very fast for large shears; in this case (for
initially overdense ellipsoids), the third-order solution does not
much improve the agreement with the numerical solution, with respect
to the second-order one.
\item Initially underdense ellipsoids can collapse if the
shear is large enough; in this case the third-order prediction is
always sufficiently accurate.
\end{enumerate}

These results can be used to predict the collapse time of a
homogeneous ellipsoid in a fast and accurate way (probably more
accurate than numerical integration). It is necessary to correct the
second- and third-order predictions in order to reproduce the correct
behaviour in the quasi-spherical overdense cases. This correction can
expressed in the following way:

\be b_c^{(nC)} = b_c^{(n)} - \Delta\exp(-ax-by)\; , \label{eq:corr} \ee

\noindent where $n$=2,3, and the three coefficients take on the
following values:

\bea      & {\rm 2nd\ ord}&\ \ {\rm 3rd\ ord} \nonumber\\
\Delta  = &\ 0.580        & {\rm or} \ \ 0.364 \\
     a  = &\ 5.4          & {\rm or} \ \ 6.5 \nonumber\\
     b  = &\ 2.3          & {\rm or} \ \ 2.8\; . \nonumber \eea

\noindent This corrections are applied only when $\delta_l>0$; no
correction is applied when $\delta_l \leq 0$. 

All these calculations have been repeated with two other background
models, namely the open $\Omega_0=0.2$ model and the flat
$\Omega_0=0.2$ model with cosmological constant. The Lagrangian
calculations are quite insensitive to the cosmology, as shown by
equations (\ref{eq:tpert}) (note that the pancake case, $x=\delta_l$
and $y=0$, being exactly predicted by Zel'dovich, is always
independent of the cosmology). As expected, the numerical calculations
give very similar results for the $b_c$ function, with slightly larger
errors when the background density becomes small and the ellipsoid
takes a long time to collapse. Fig. 1d shows the $\Omega=0.2$ case; an
initial overdensity $\delta_l=3$ has been chosen to allow the collapse
of all the ellipsoids reproduced in the figure. Then, the
above-mentioned results for $b_c$ can be used in any of the
cosmologies checked here.

In conclusion, the Lagrangian series can be used to accurately
approximate the collapse of a homogeneous ellipsoid. On the other
hand, homogeneous ellipsoid collapse can be seen as a particular
truncation of the Lagrangian series. Its ability to reproduce the
collapse time of generic perturbations will be shown in the next
section.  In this case, ellipsoidal collapse has to be considered not
as a description of the dynamics (or even the geometry!) of an {\it
extended} region, but as an approximate description of the {\it local}
dynamics of a mass element.

Before going on, it is useful to comment on the concept of `locality'.
From a mathematical point of view, the evolution of a continuous
system is local if its evolution equations are ordinary differential
equations, i.e. with no partial derivatives\footnote{It is assumed
that there is no explicit coupling between field values at different
points}. Thus, any point's trajectory is fully determined by its
initial conditions, i.e. it is not necessary to evolve all the
trajectories together. From a physical point of view, non-locality can
be associated to the long-range character of gravitational forces: the
fate of a mass element depends on all the other mass elements. The
non-local character of gravitational dynamics has recently been
stressed again in a paper by Kofman \& Pogosyan (1995). Nonetheless,
some dynamical approximations predict `local' dynamical evolution. The
unrealistic spherical model is such a one: the fate of a spherical
perturbation depends only on its initial density. At variance, the
Zel'dovich approximation gives some information on non-local tidal
forces: the motion of a mass element depends on all other
elements. This information is contained in the gravitational
potential, which is related to the density by a non-local Poisson
equation. Thus the Zel'dovich approximation is physically
non-local. But, once the initial peculiar potential is known, the
dynamical evolution is local, in the mathematical sense given
above. In other words, the initial conditions contain non-local
information, while the evolution is local. The same holds true for the
homogeneous ellipsoidal collapse model.  Analogous considerations can
be applied to the Lagrangian perturbation theory at any order, as the
equations for the perturbative terms are separable in space and time,
as recently demonstrated by Ehlers \& Buchert (1996). If this is the
case, the non-local space equations (\ref{eq:spert}) give non-local
initial conditions, while the subsequent dynamics is independently
determined for every single trajectory.  But, while it is relatively
simple to obtain the statistical distribution of the initial
conditions for the Zel'dovich approximation, namely of the $\lambda_i$
eigenvalues, the analytical determination of the statistical
distribution of other contributions leads to discouraging mathematical
difficulties. Nonetheless, as a consequence of the mathematical
`locality' of Lagrangian perturbations, the perturbative terms have to
be calculated just once at the beginning, making this kind of
calculation dramatically faster than usual N-body simulations.

%%%%%%%%%%%%%%%%%%%%%%%%%%%%%%  5  %%%%%%%%%%%%%%%%%%%%%%%%%%%%%%%%

\section{Collapse time in the Gaussian field case}

In this section, Gaussian fields with scale-free power spectra are
considered. To proceed analytically, the equation $J=\det(\delta_{ab}
+S_{a,b})=0$ ought to be solved for general initial potentials, and
the probability distribution function (hereafter PDF) of its smallest
non-negative root ought to be obtained. As a matter of fact, it is
very hard to get the whole PDF of the various contributions to the
deformation tensor. Even neglecting the non diagonalizable 3c term,
the various contributions to $S_{a,b}$ are not diagonal in the same
frame. It is definitely convenient to get the PDF of the collapse
times by constructing realizations of Gaussian fields in cubic grids.

In these calculations the grid does not need to be very large: what is
needed is a sufficient degree of non-locality, which is provided even
by small grids. 16$^3$ and 32$^3$ grids have been used, with identical
results.  Initial potentials have been simulated, following what is
done for initial conditions of N-body simulations. Power spectra have
been chosen as $P_\varphi(k) \propto k^{n-4}$, with $n$=--2, --1, 0
and 1; these correspond to matter perturbation spectra $P_\delta(k)
\propto k^{n}$. Spectra have been normalized with $\sigma=1$, where
$\sigma$ is the total density variance. Of course, any other
normalization can be obtained by a time rescaling.
 
Ten realizations have been performed for every power spectrum. Poisson
equations (\ref{eq:spert}) have been solved with Fast Fourier
Transform (FFT) techniques; FFT have also been used to calculate the
derivatives of the various potentials. The growing mode $b$ has been
used as a time variable, and the $\Omega$ dependence of the time
functions $b_n$ (equations \ref{eq:tpert}) has again been neglected.
The following collapse time estimates have been calculated for every
point: spherical collapse (hereafter SPH), 1st-order or Zel'dovich
approximation (1ST), 2nd-order (2ND), 3rd-order (3RD) and ellipsoidal
collapse (ELL). SPH, 1ST and ELL collapse times have been calculated
analytically, on the basis of the $\lambda$ eigenvalues of
$\varphi_{,ab}$ at any point; ELL has been calculated at third-order
and corrected around the spherical value as in equation
(\ref{eq:corr}). 2ND and 3RD have been calculated by looking for the
instant at which $J<0$, then using conventional root-finding
algorithms. It is possible that $J$ becomes negative and then positive
soon after; these events can be lost if the search is not fine
enough. As a matter of fact, a very small number of points, on the
order of a few times 10$^{-5}$ of the total number, were missed by the
searching algorithm.

\begin{figure*}
\centerline{
\psfig{figure=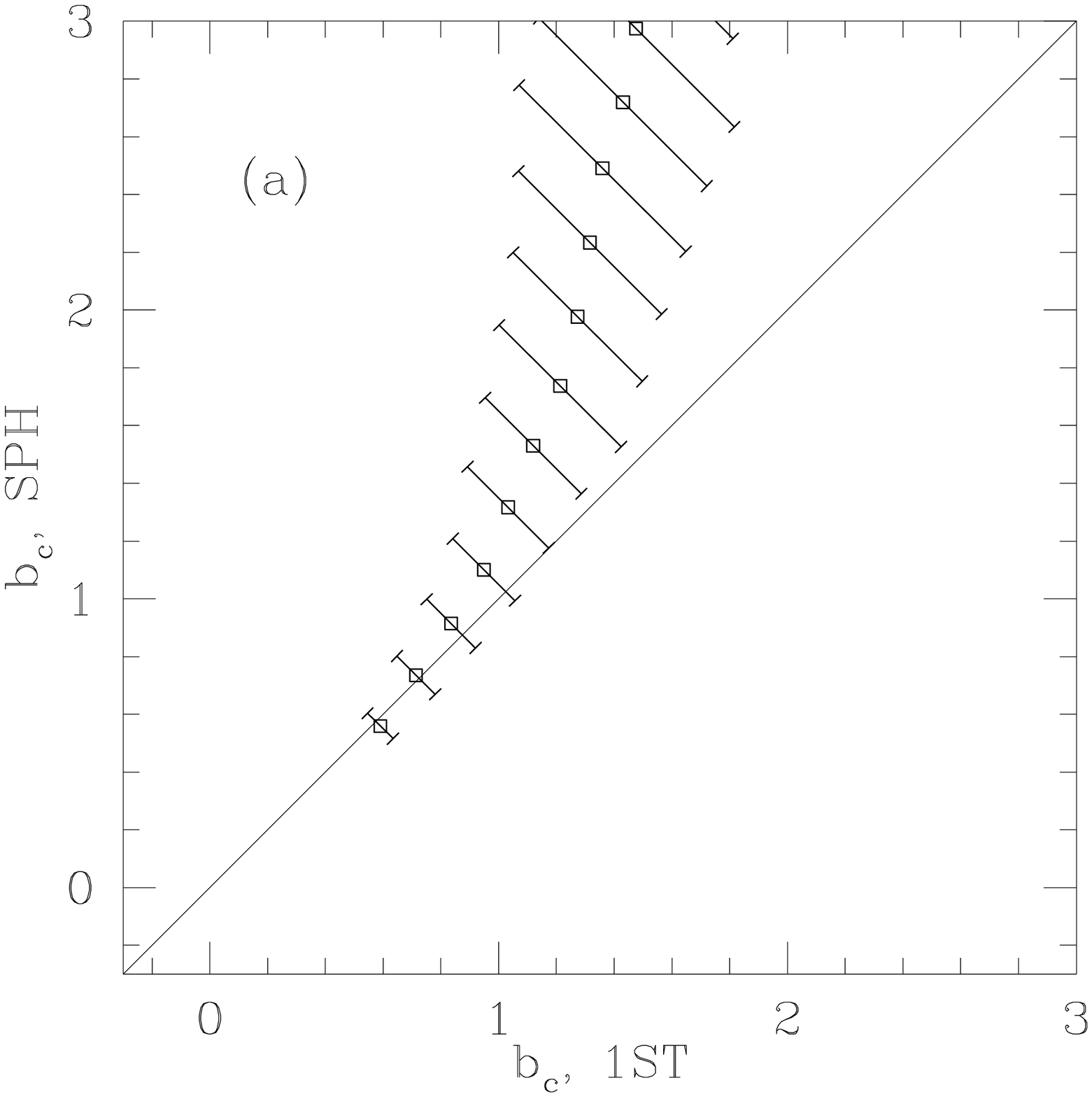,width=5cm}
\psfig{figure=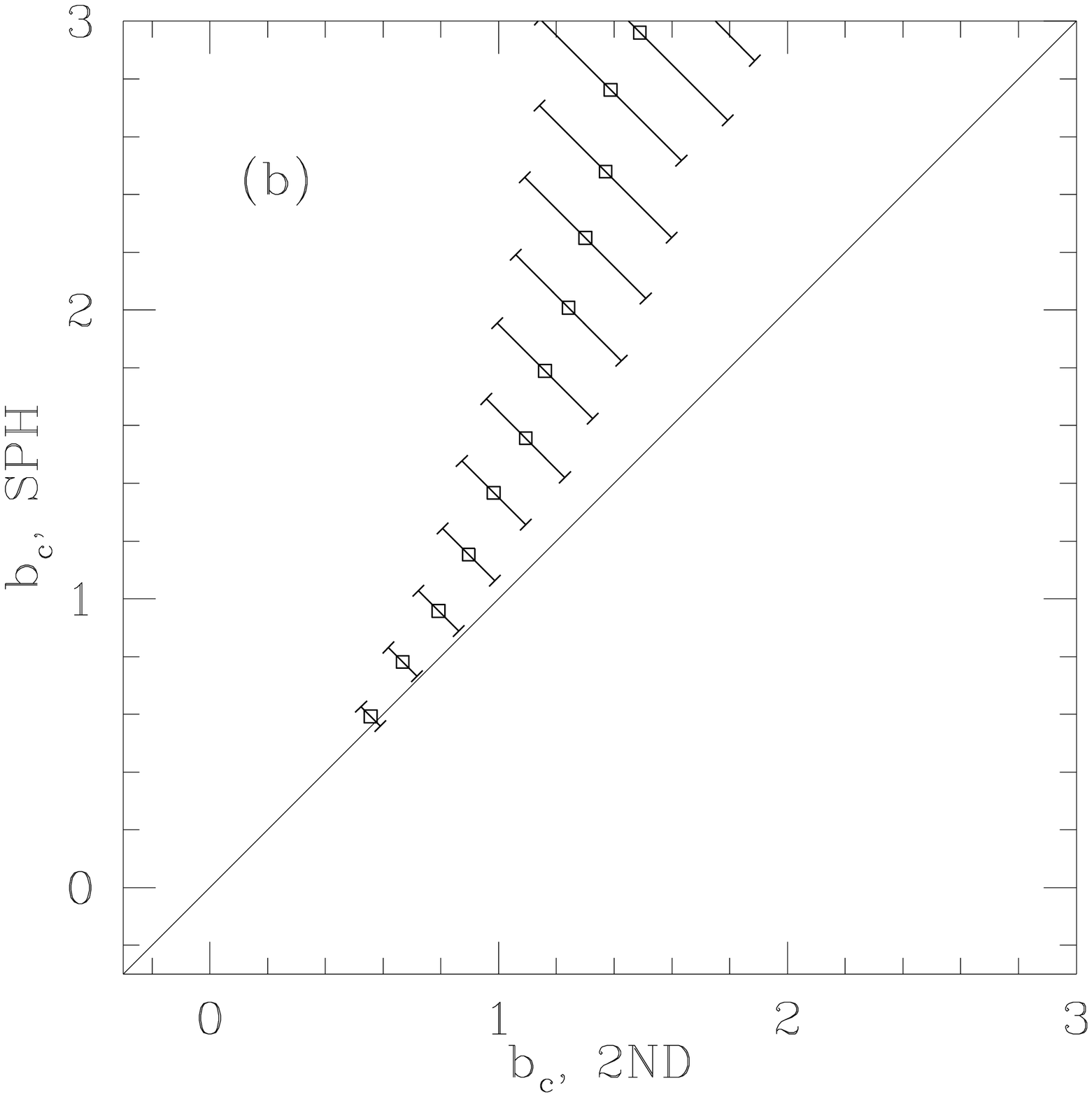,width=5cm}
\psfig{figure=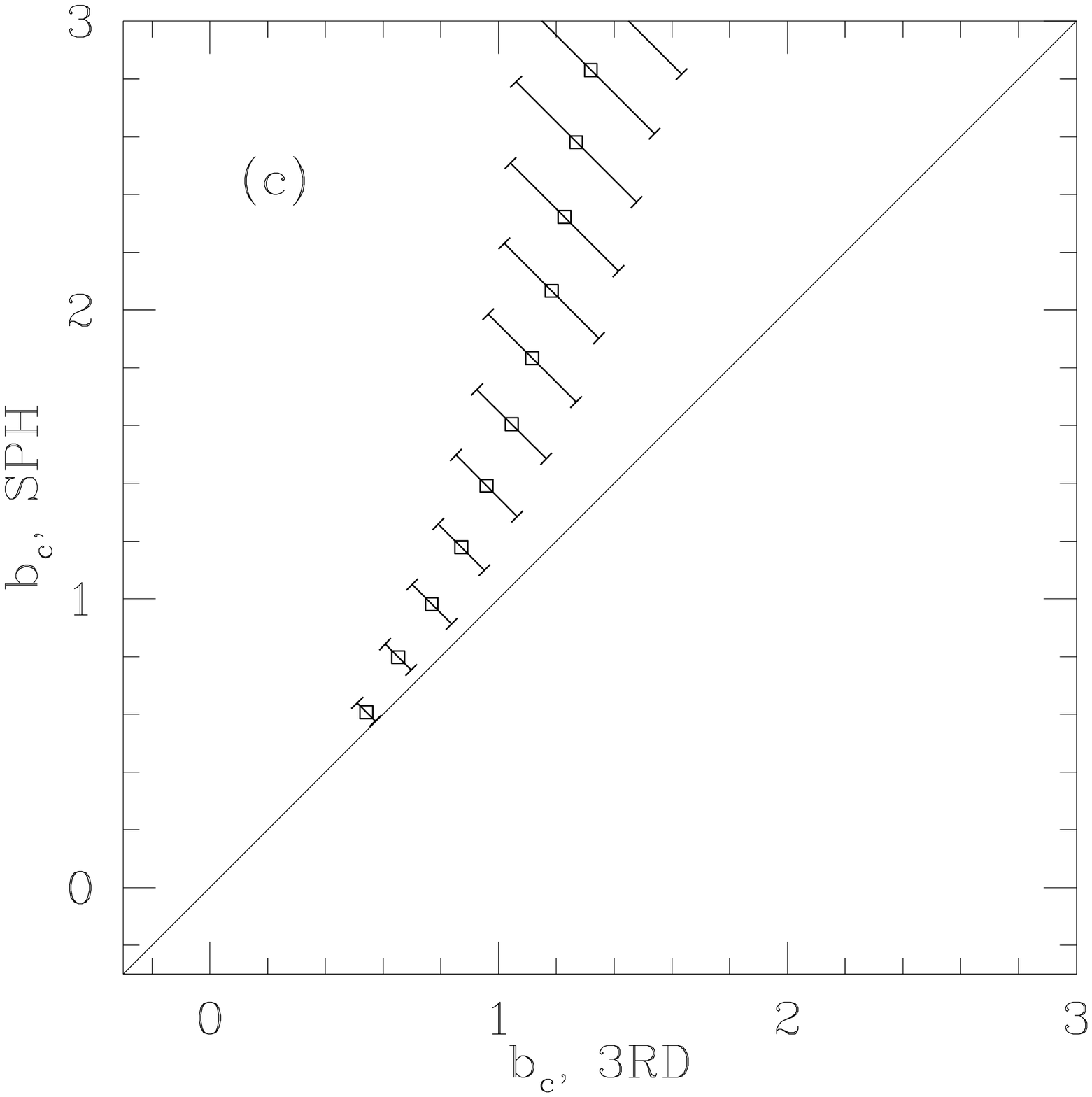,width=5cm}
}
\centerline{
\psfig{figure=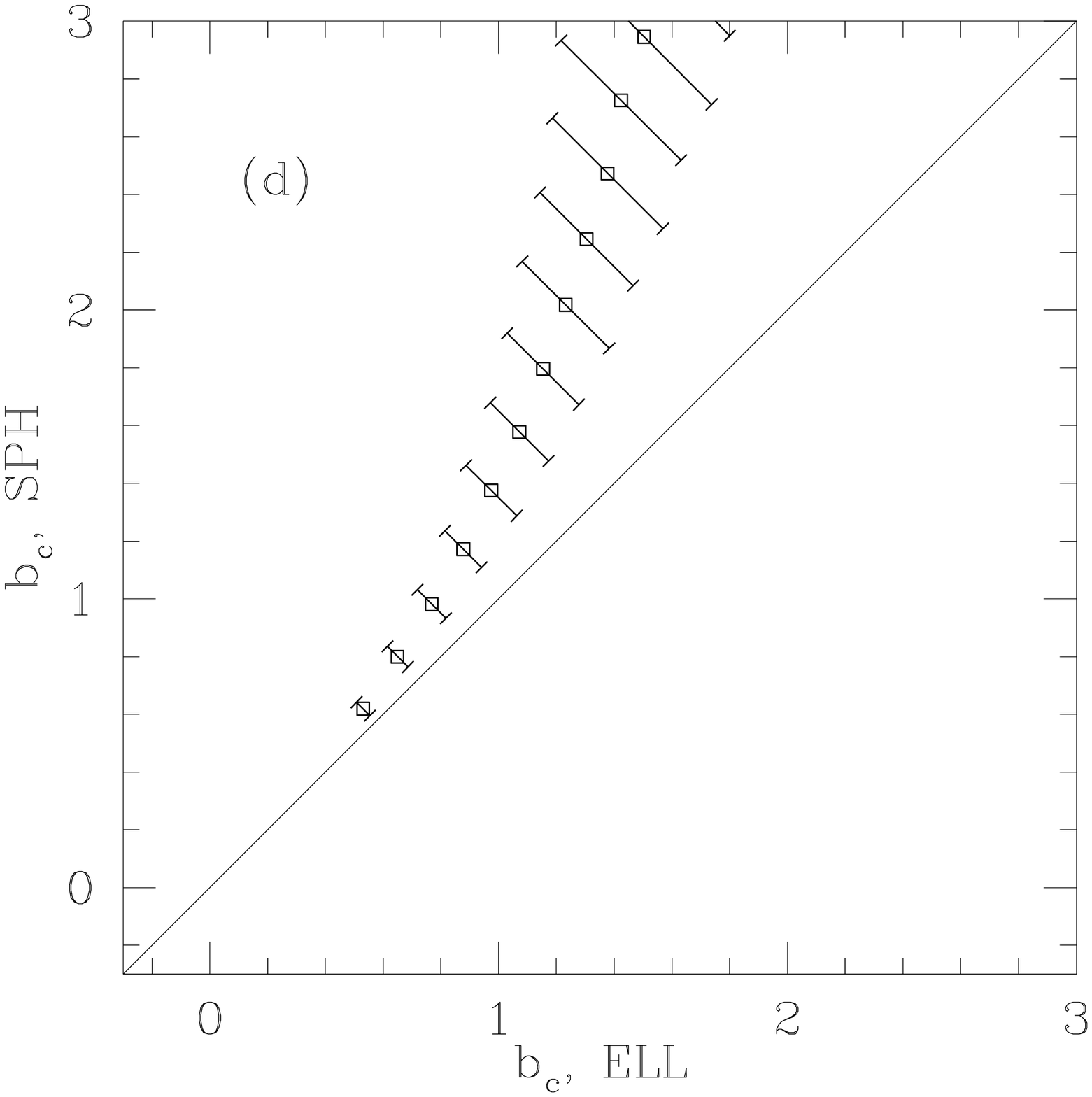,width=5cm}
\psfig{figure=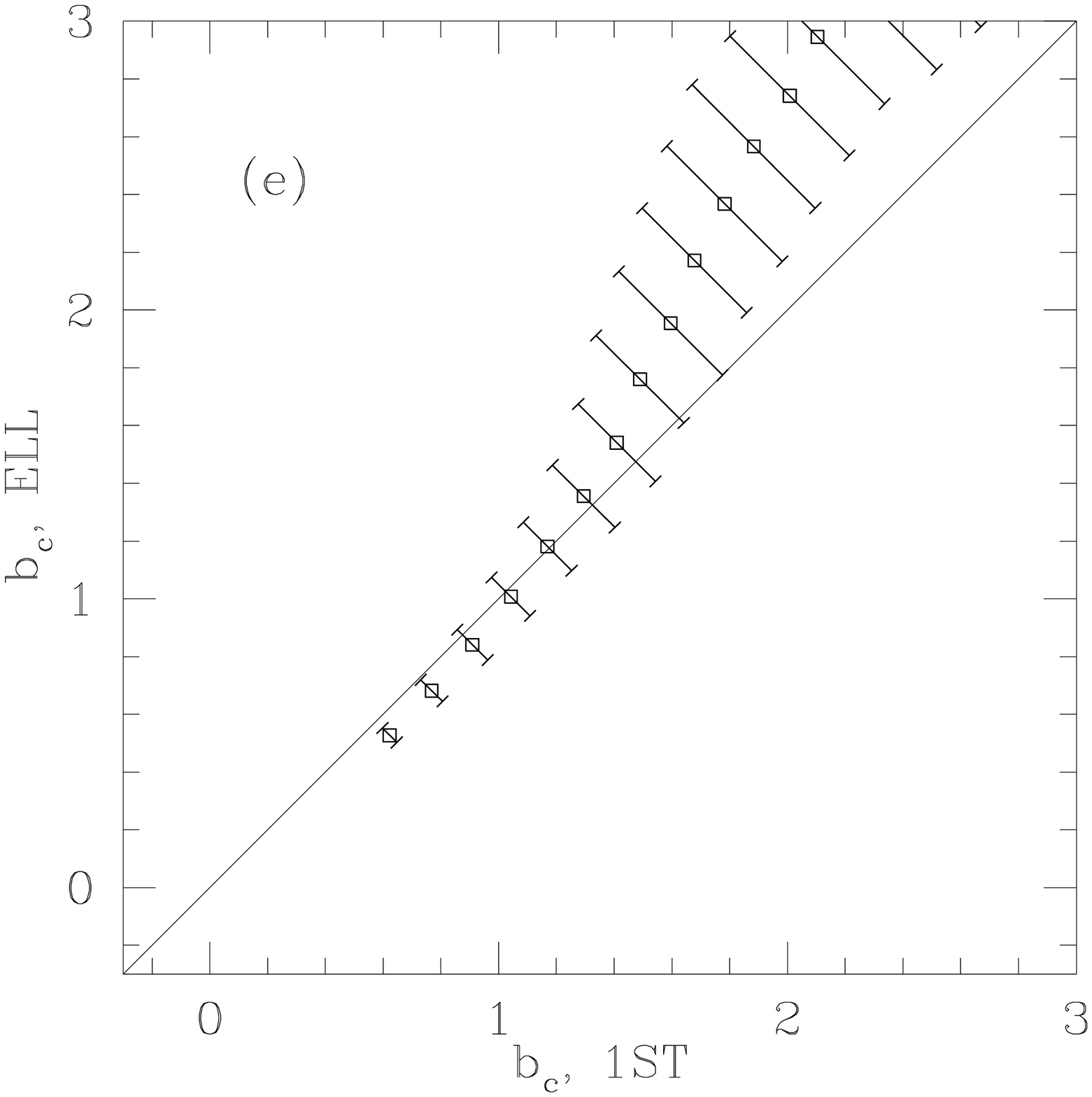,width=5cm}
\psfig{figure=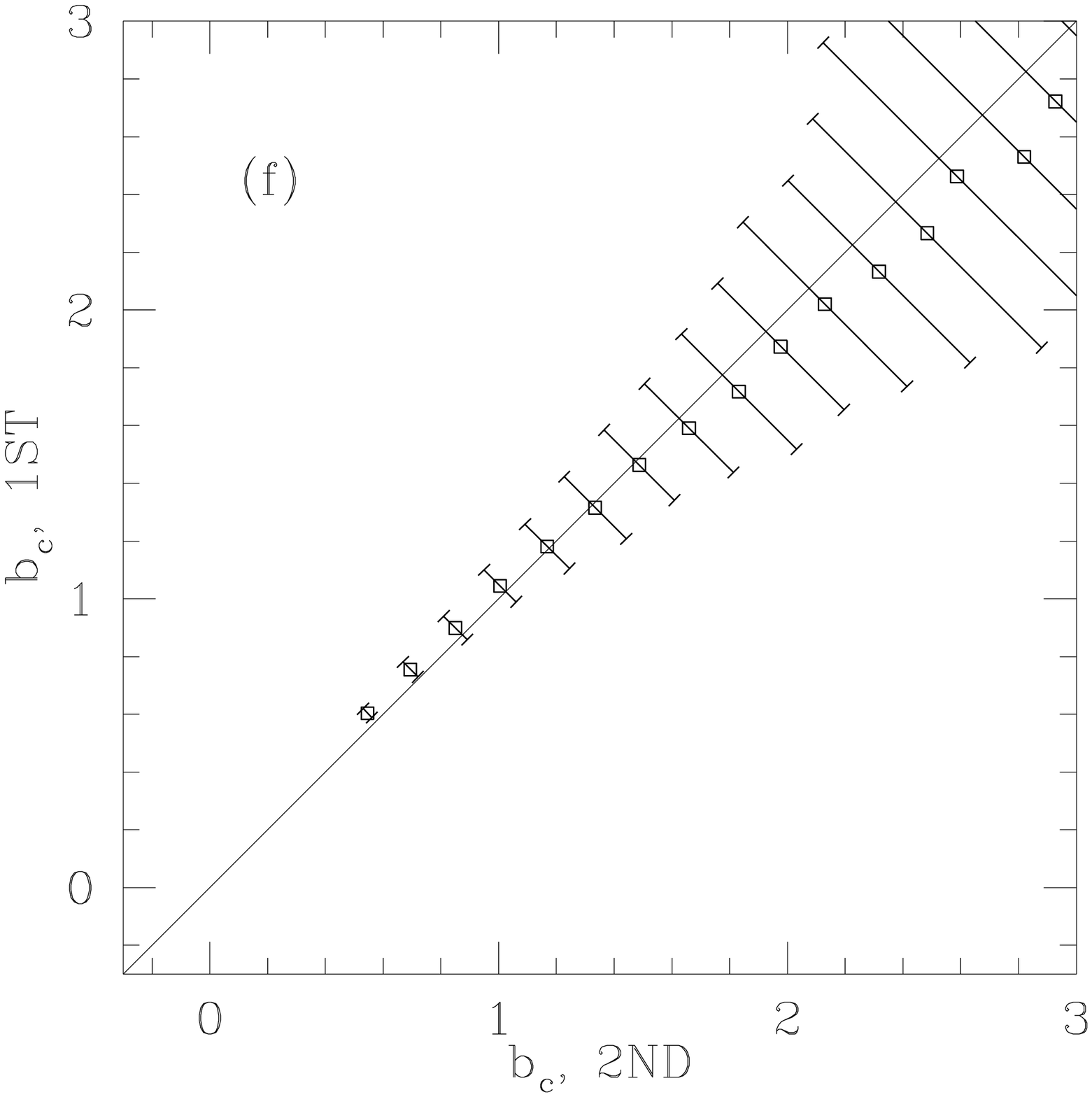,width=5cm}
}
\centerline{
\psfig{figure=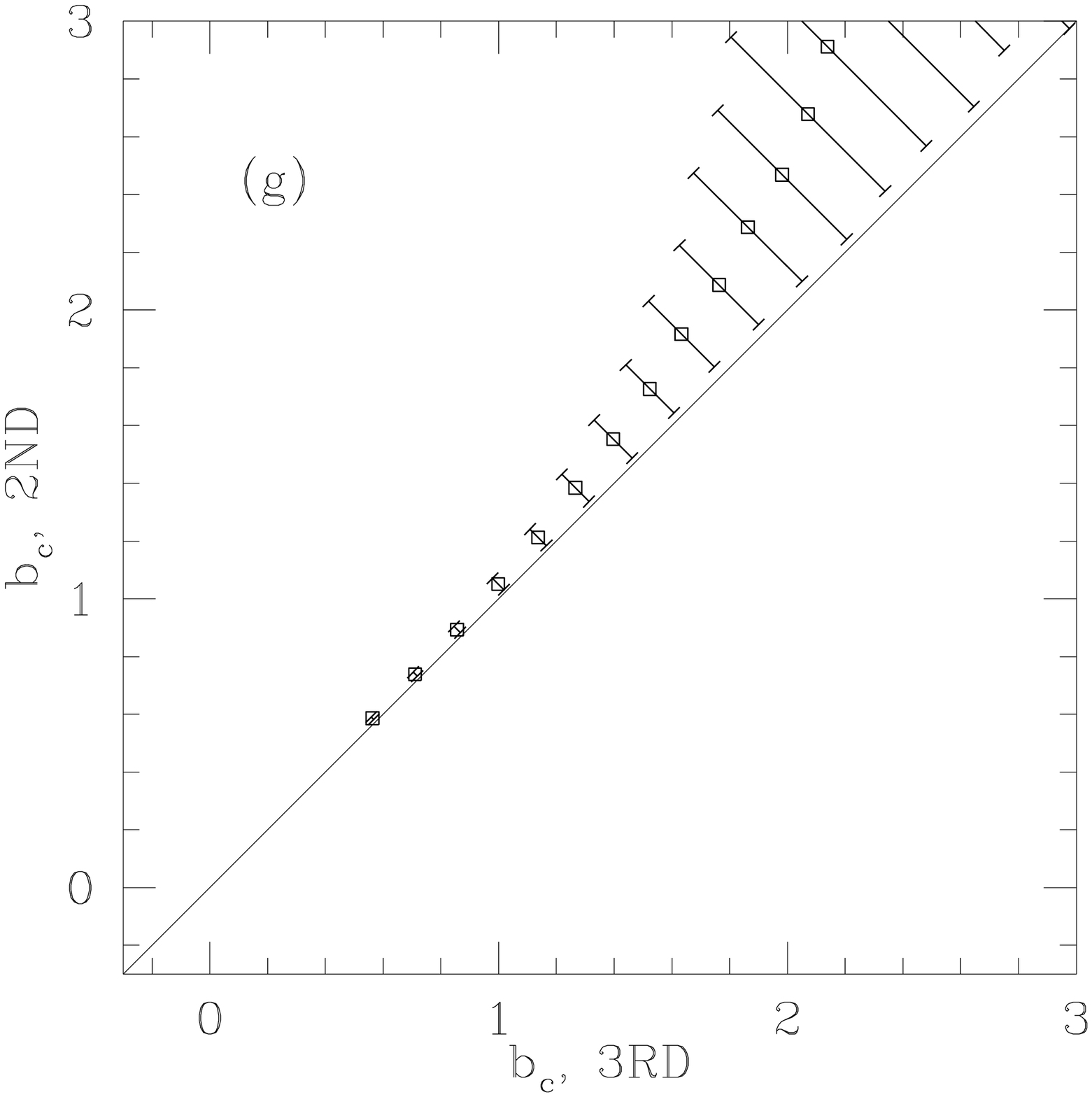,width=5cm}
\psfig{figure=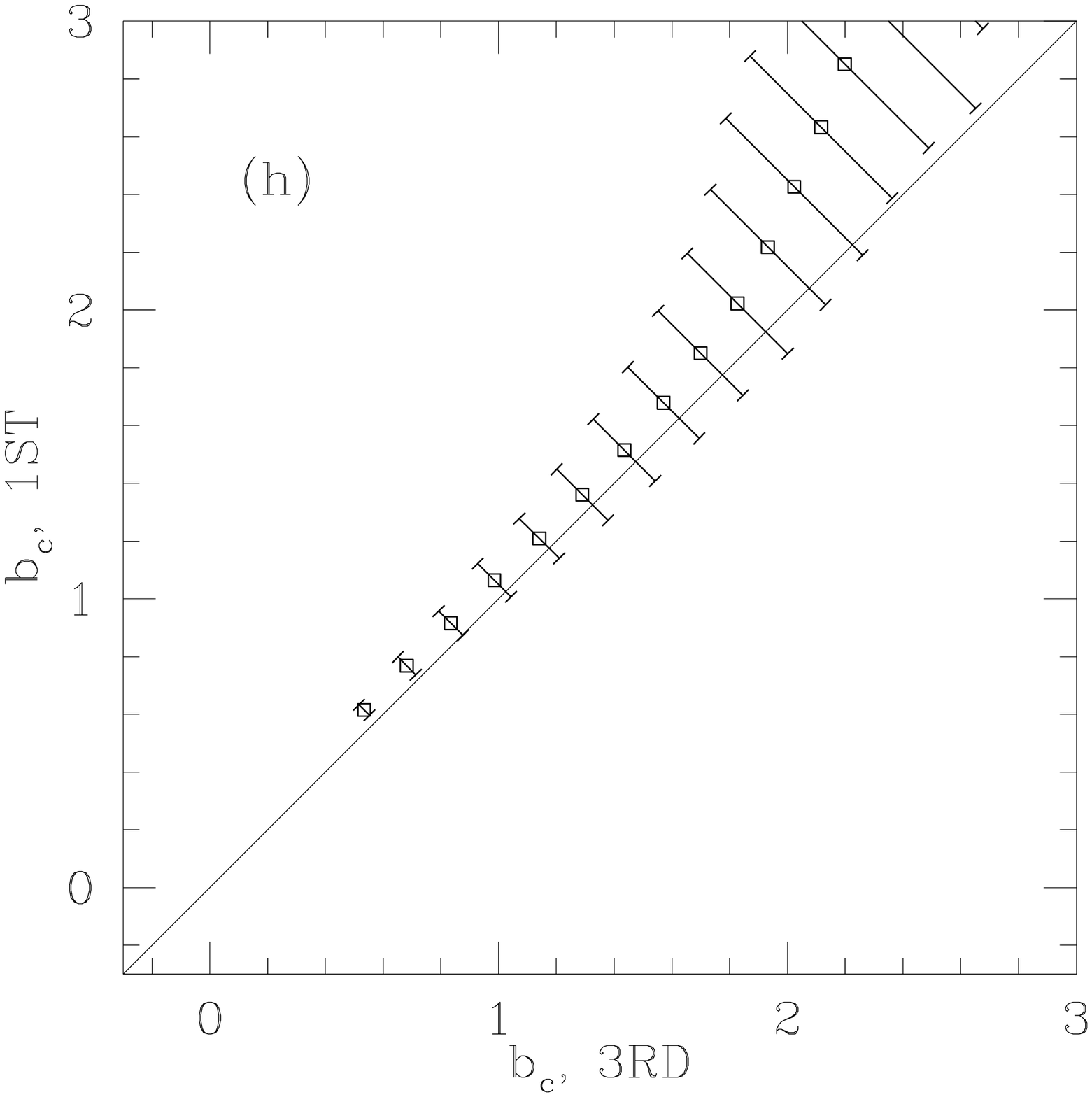,width=5cm}
\psfig{figure=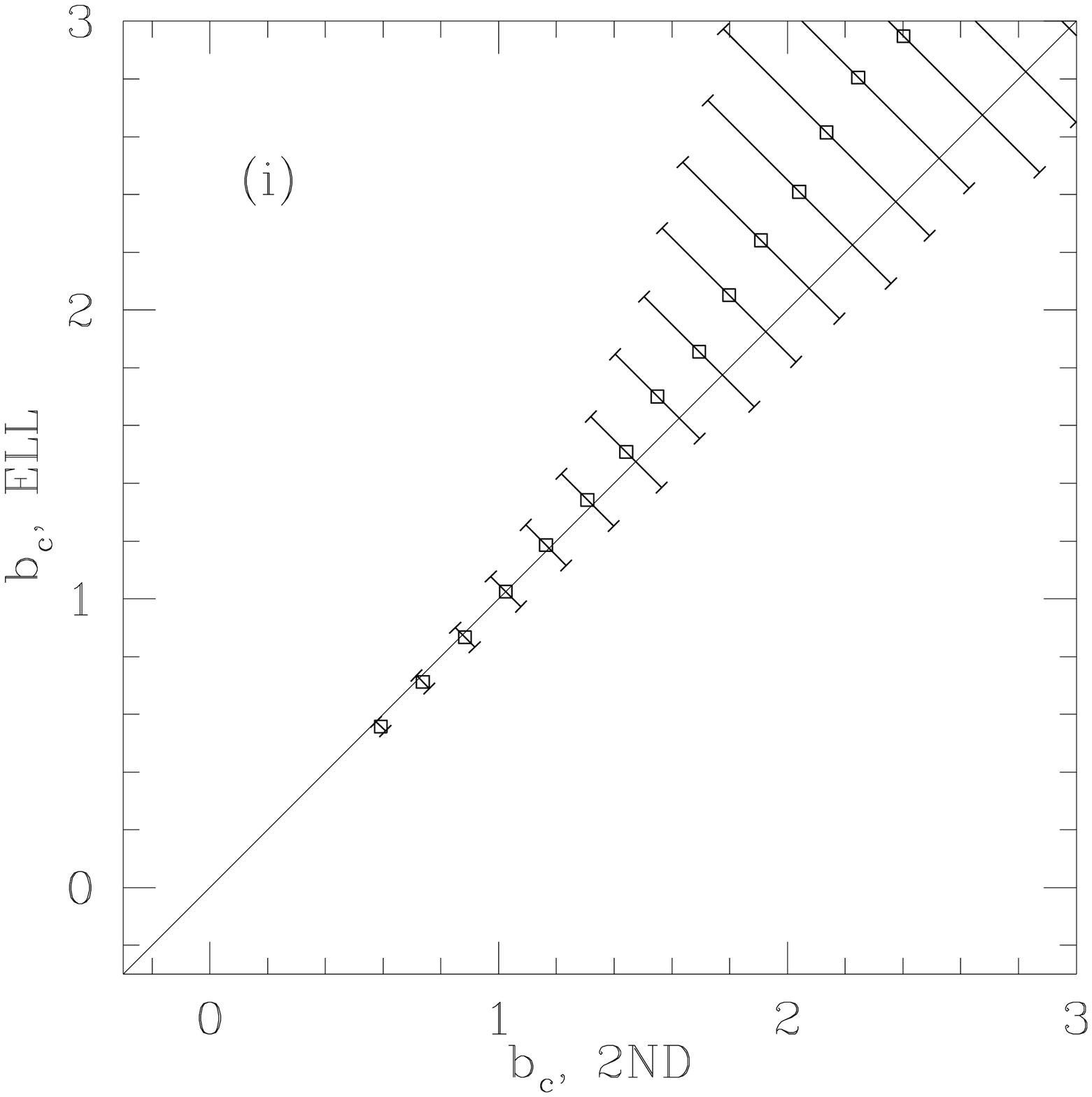,width=5cm}
}
\centerline{
\psfig{figure=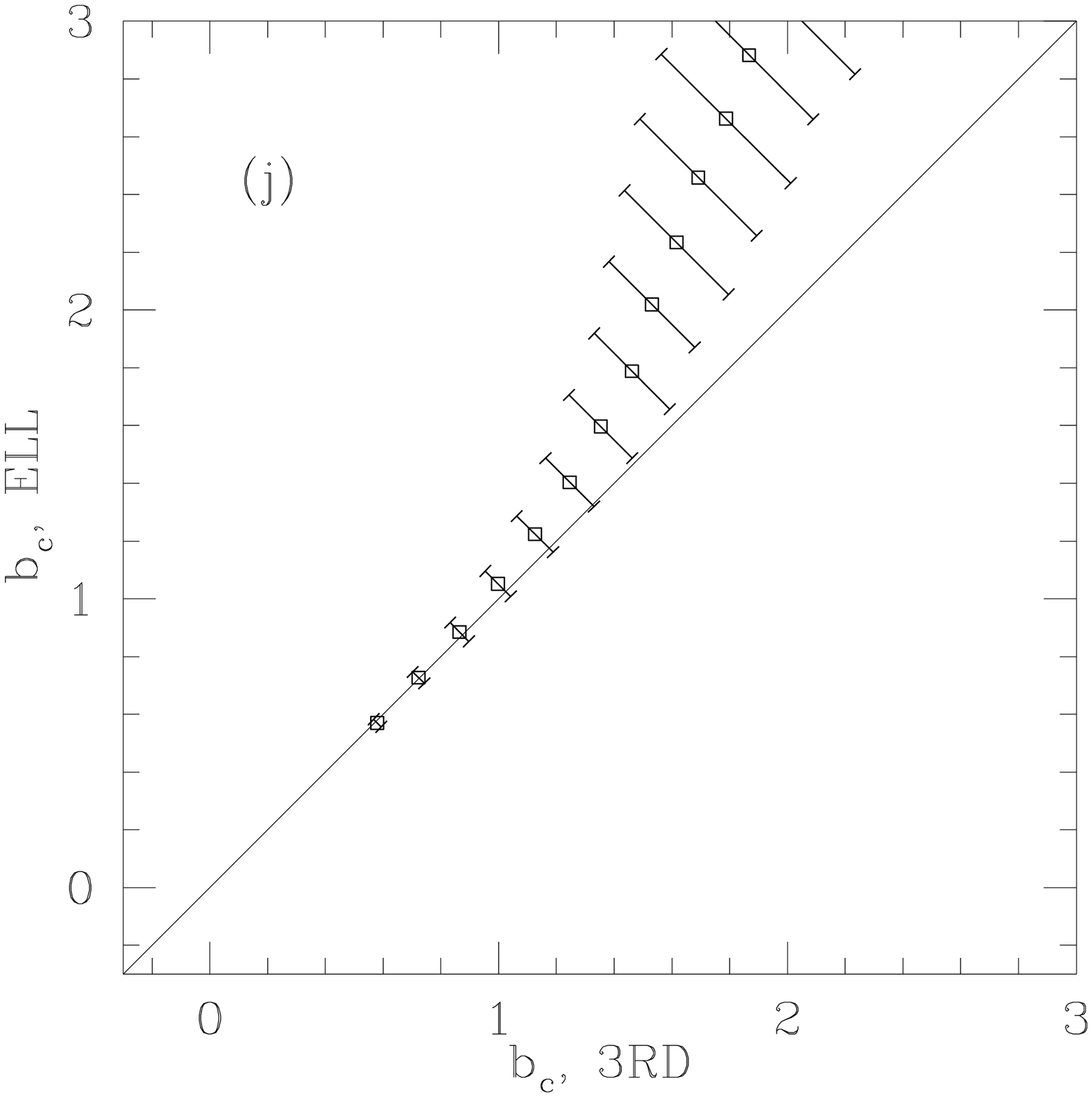,width=5cm}
}
\caption{Scattergrams of the various collapse time
estimates $b_c$, for a Gaussian field with $n=-2$.}
\end{figure*}

\begin{figure*}
\centerline{
\psfig{figure=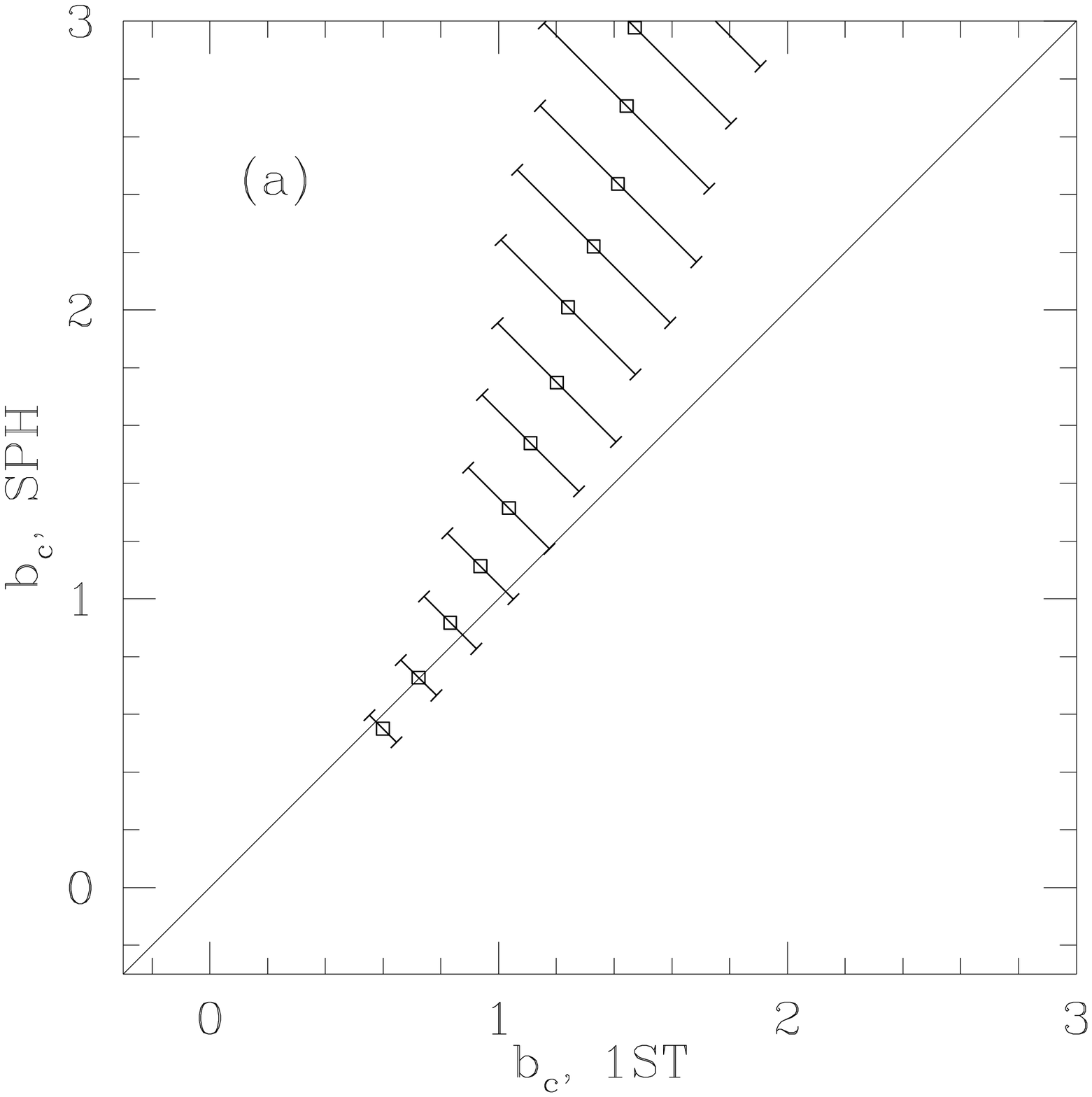,width=5cm}
\psfig{figure=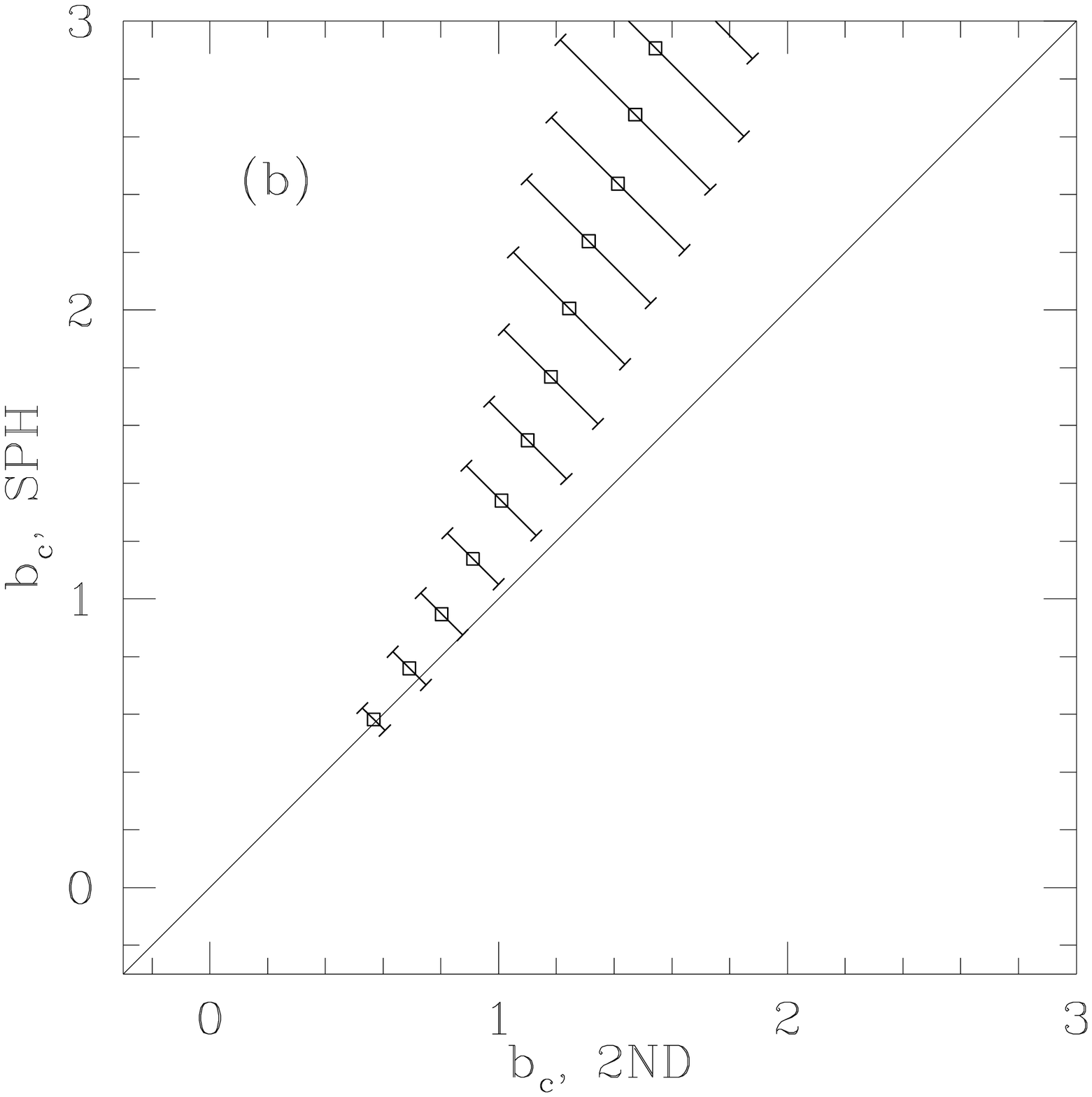,width=5cm}
\psfig{figure=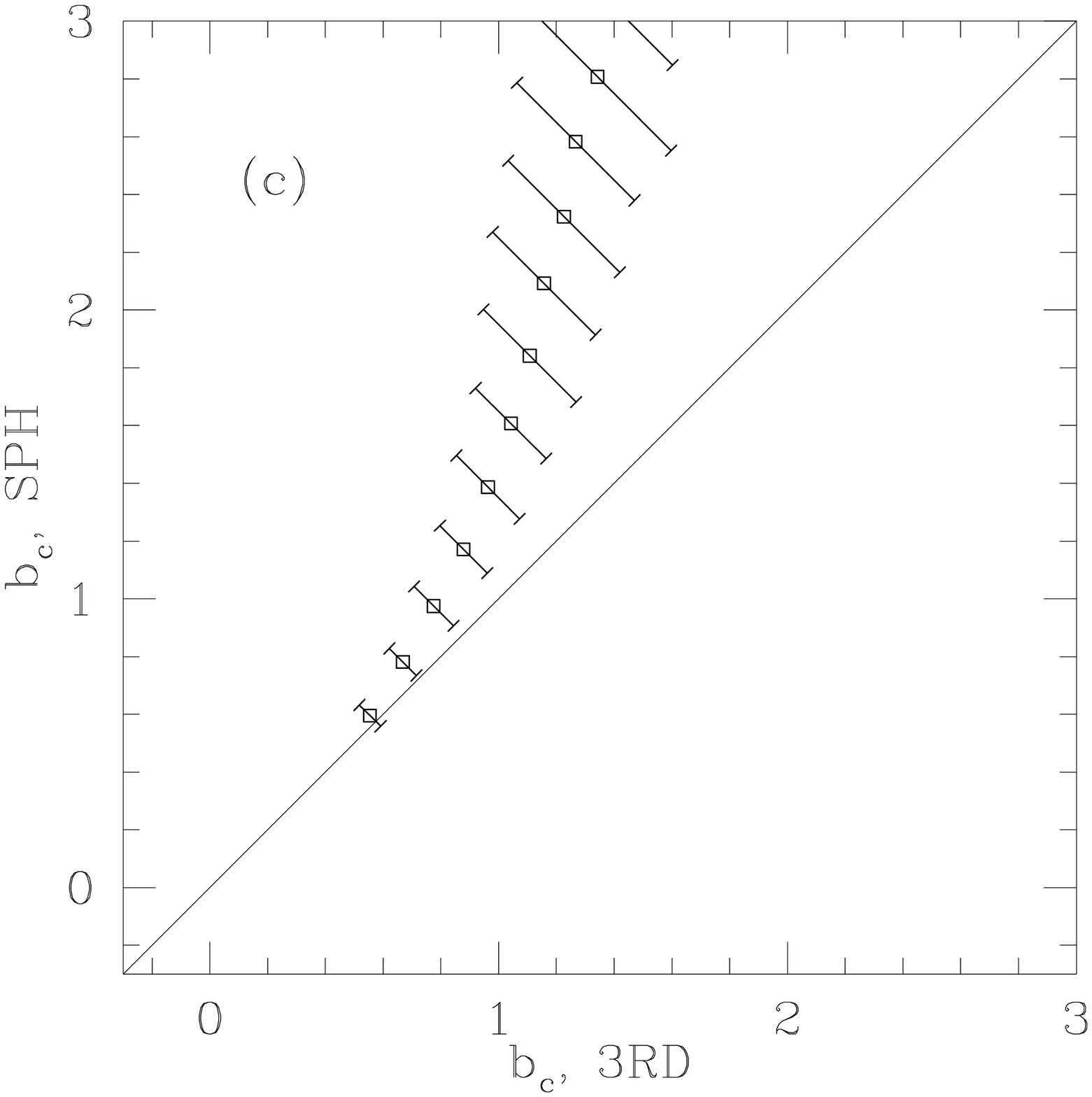,width=5cm}
}
\centerline{
\psfig{figure=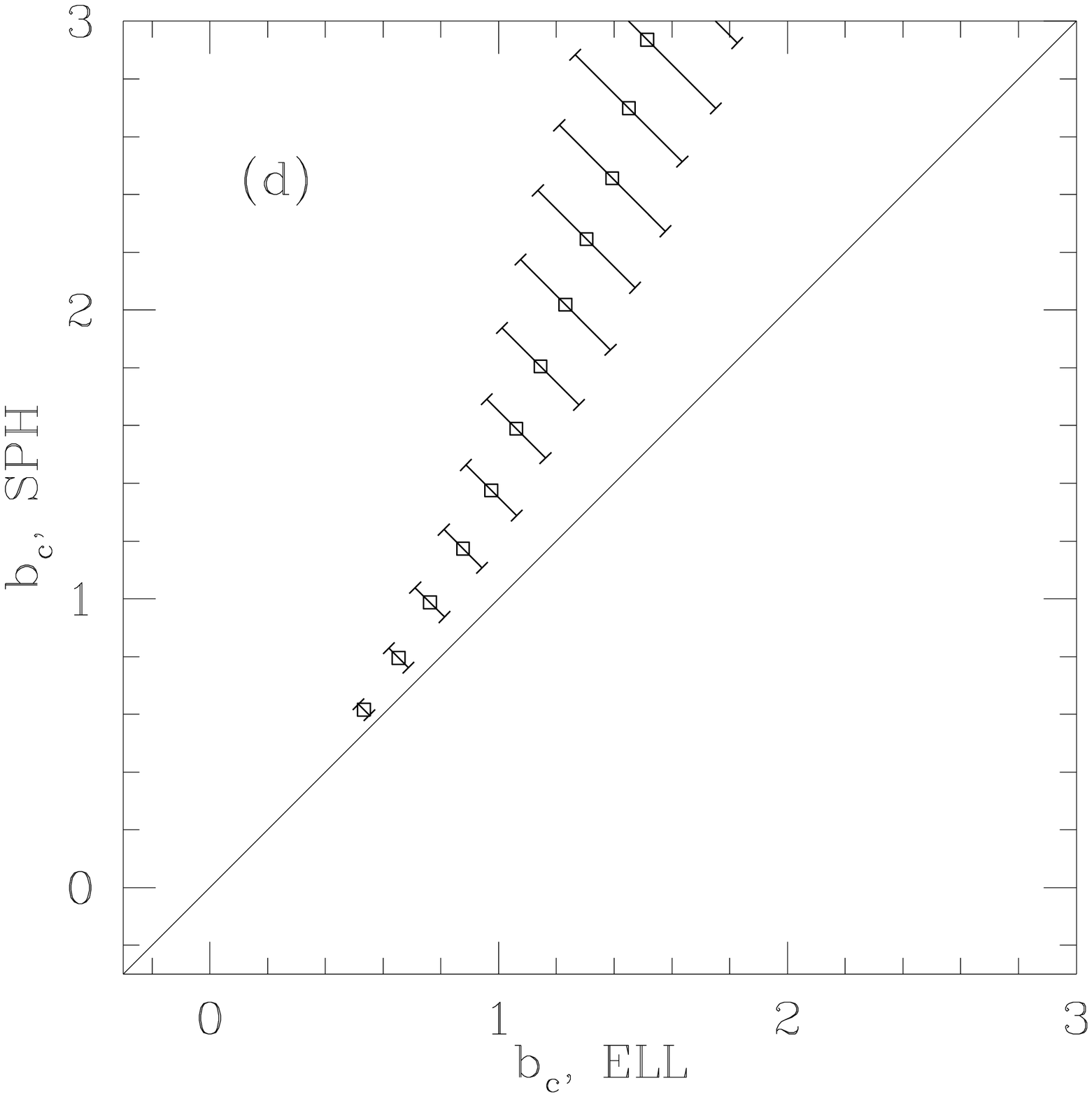,width=5cm}
\psfig{figure=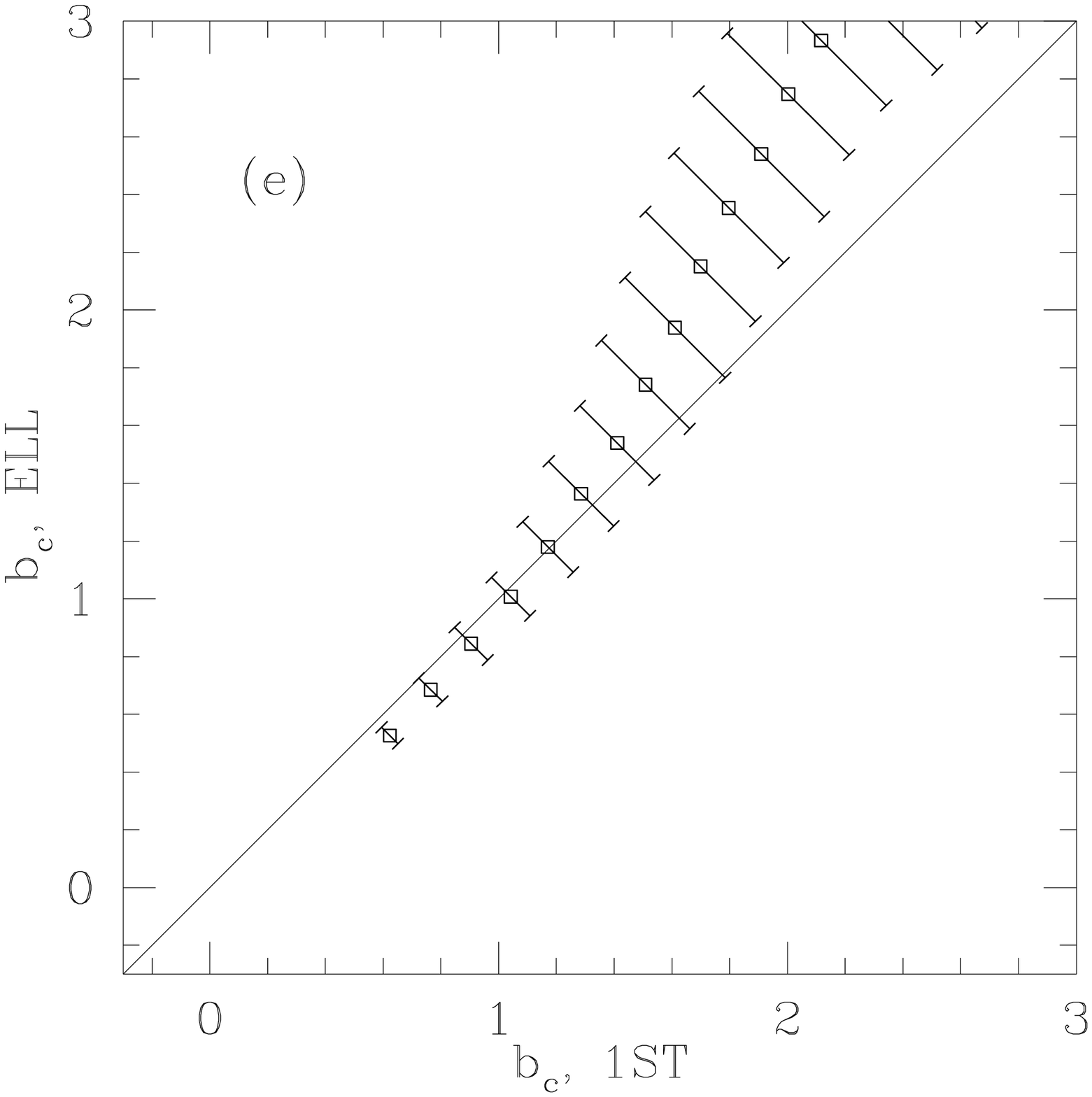,width=5cm}
\psfig{figure=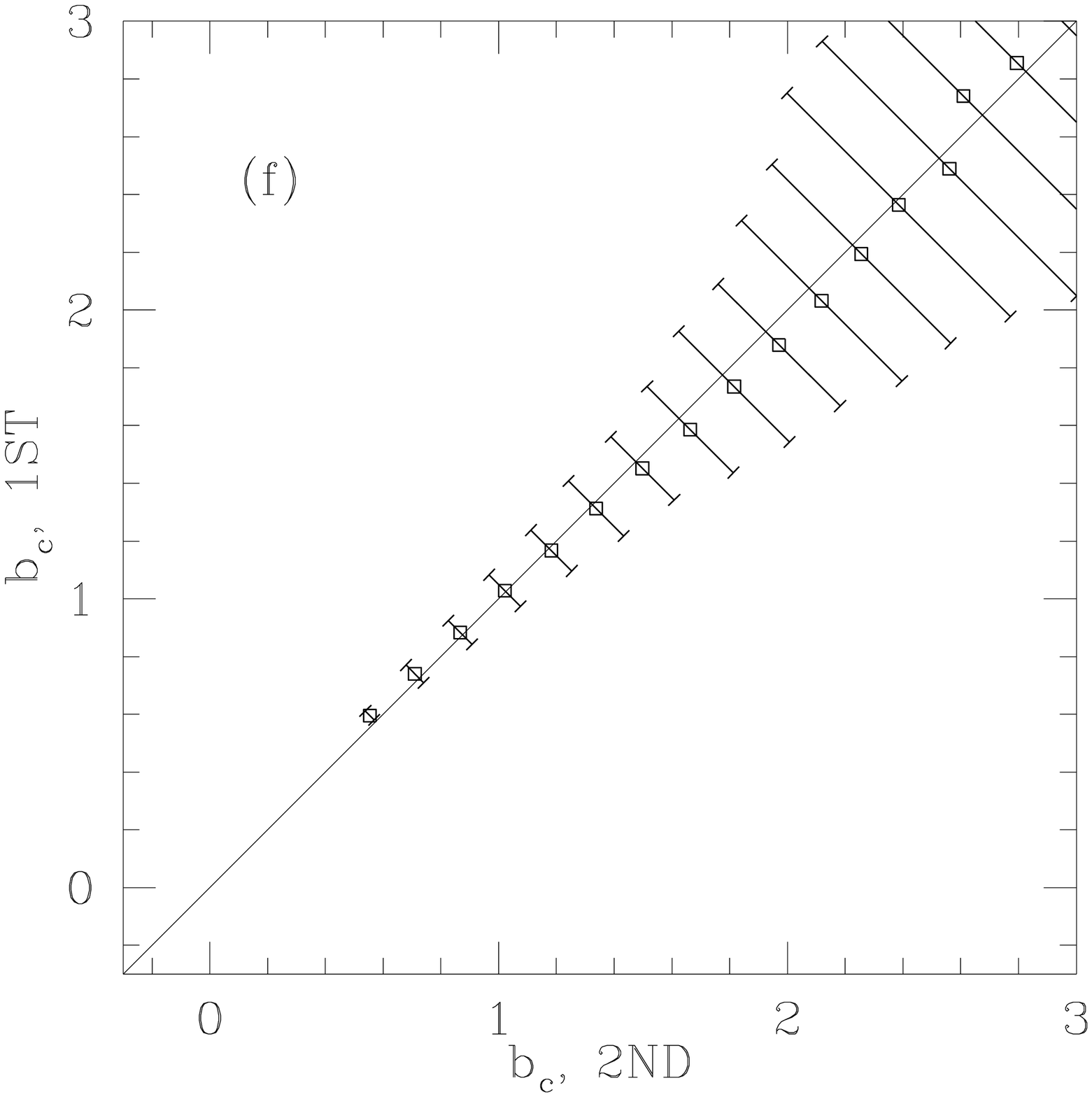,width=5cm}
}
\centerline{
\psfig{figure=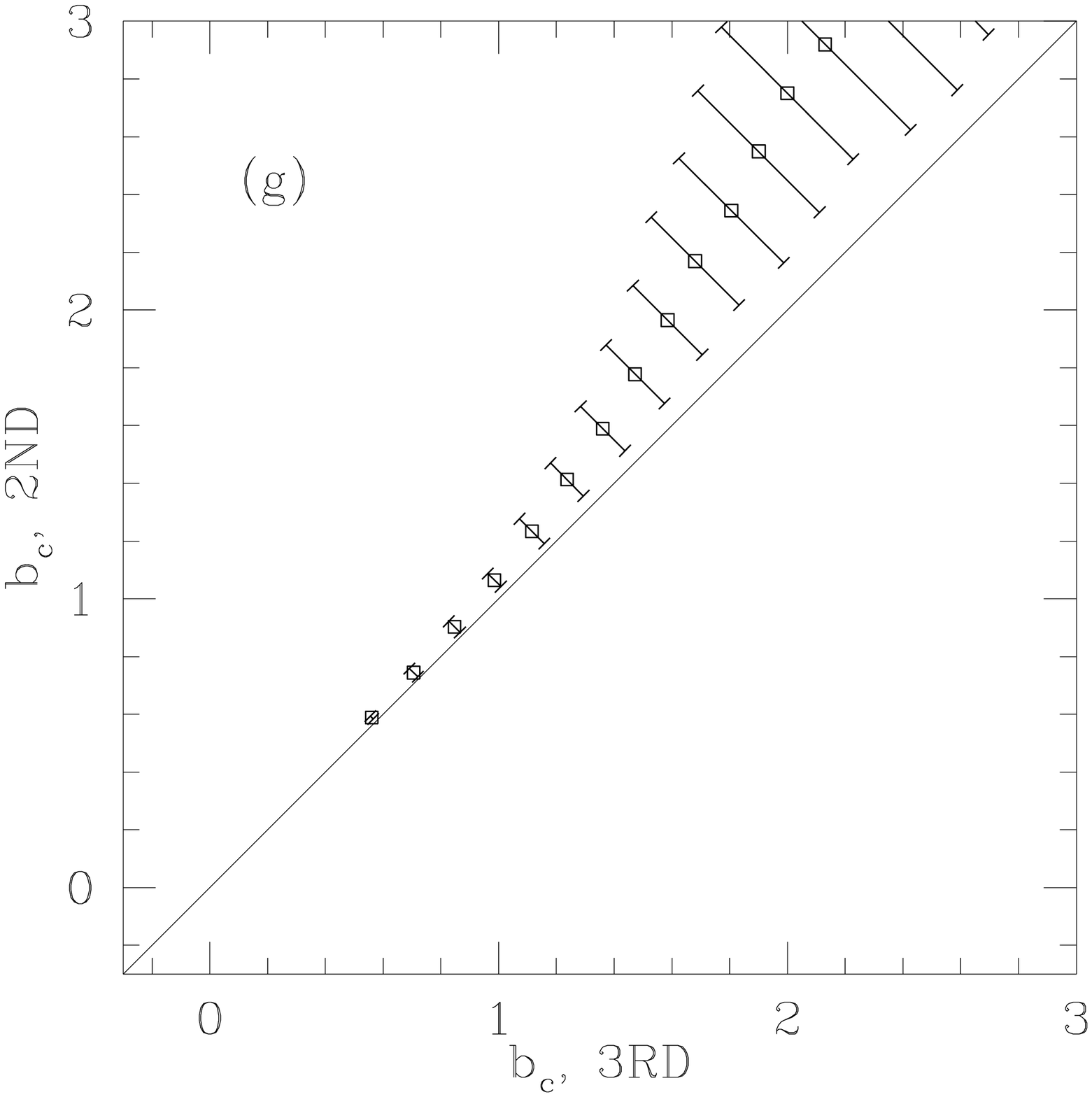,width=5cm}
\psfig{figure=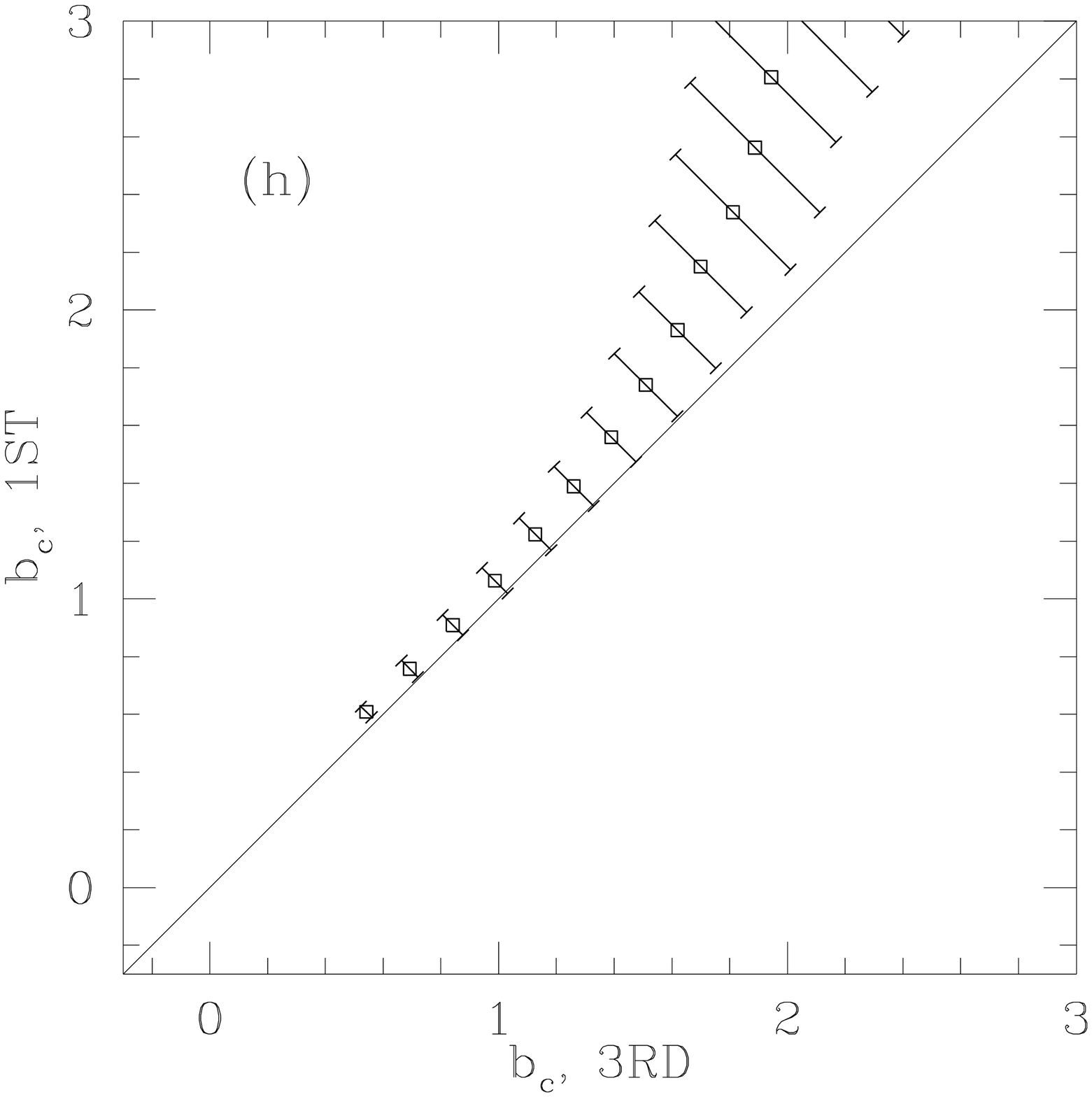,width=5cm}
\psfig{figure=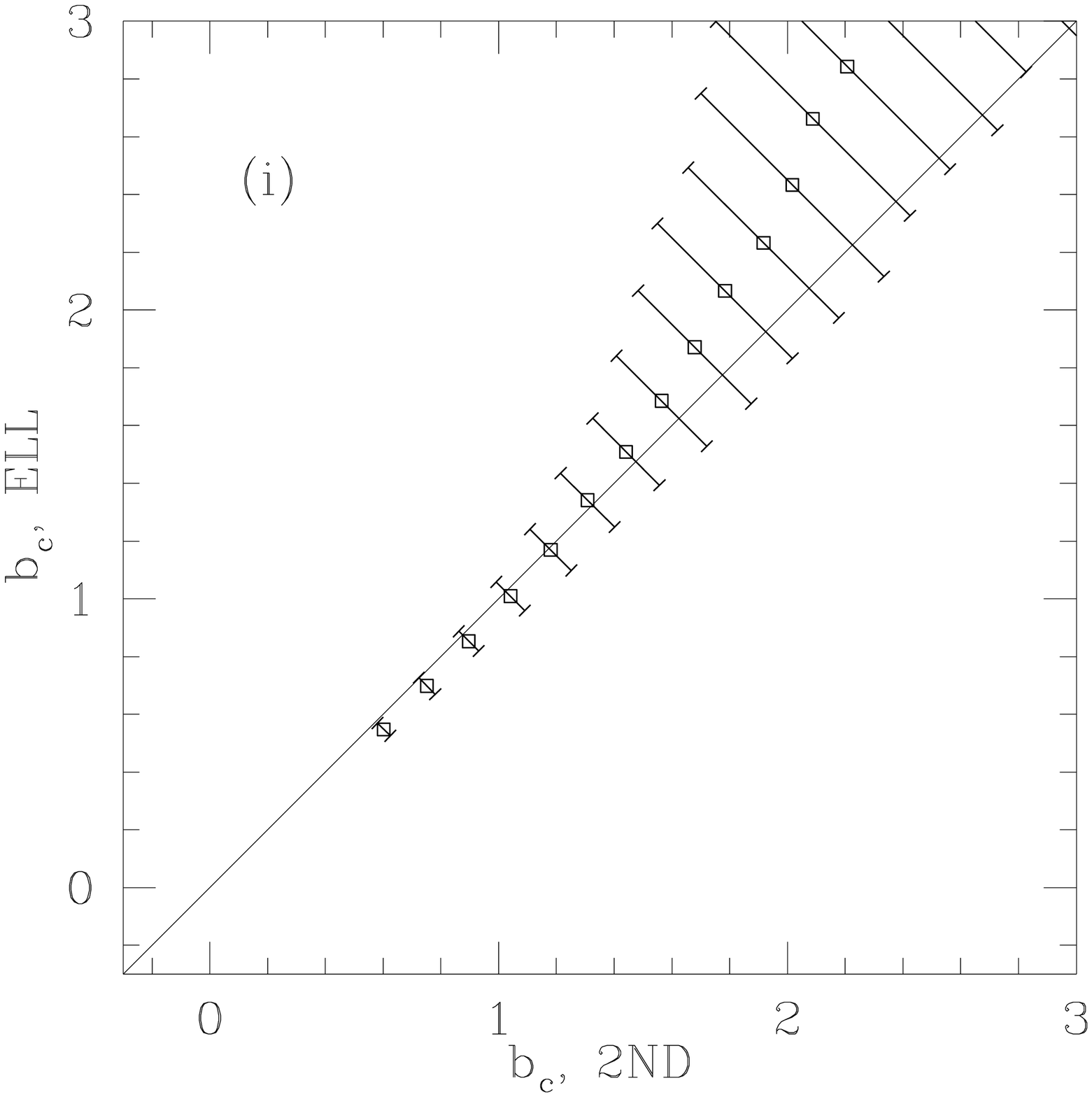,width=5cm}
}
\centerline{
\psfig{figure=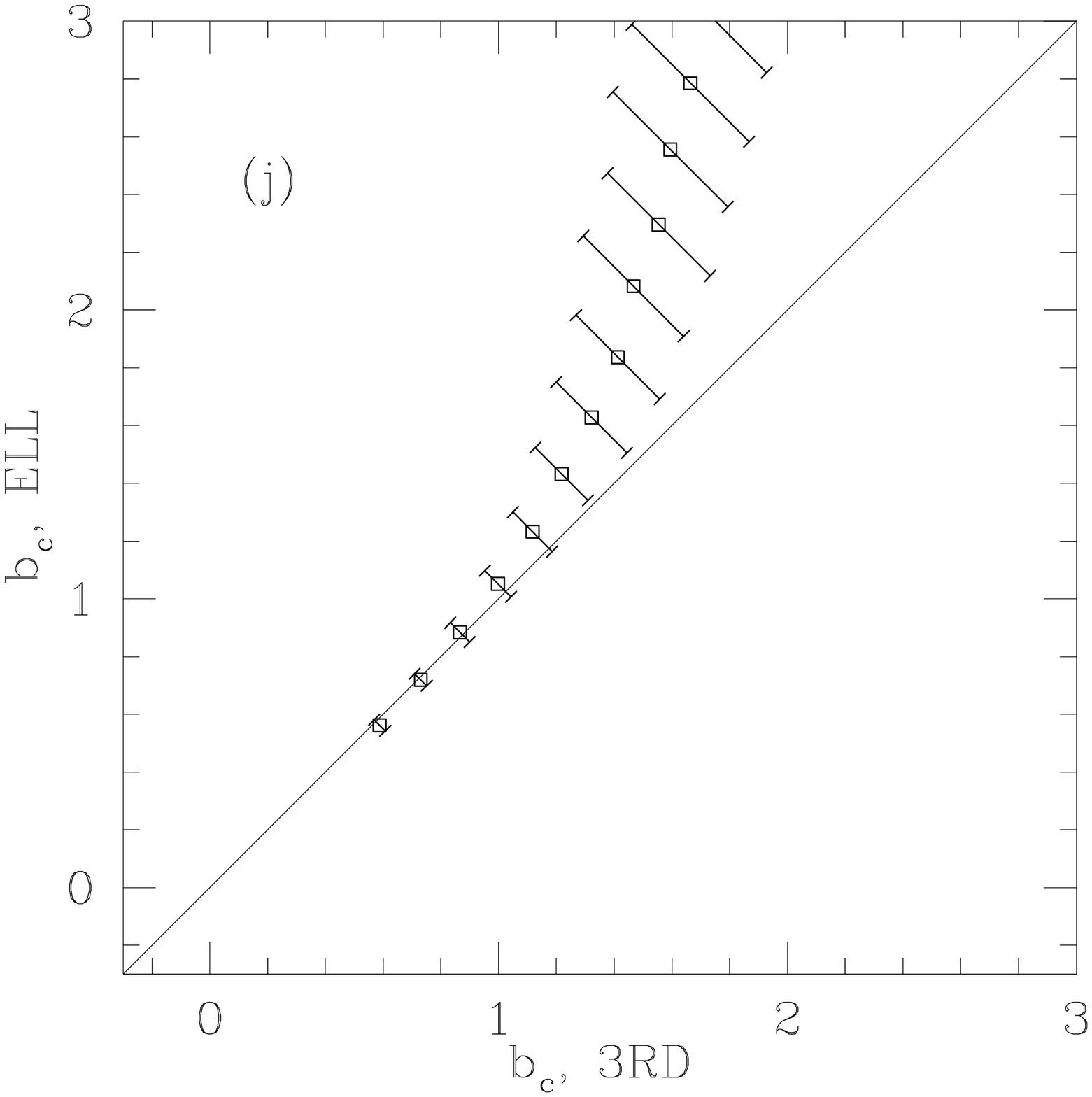,width=5cm}
}
\caption{Scattergrams of the various collapse time
estimates $b_c$, for a Gaussian field with $n=1$.}
\end{figure*}

Buchert, Melott \& Wei\ss\ (1994) state that third-order terms are very
sensitive to numerical calculations.  The same thing has been found in
these analyses; third-order calculations are not expected to be very
precise. This has been noted especially for the transversal 3c
contribution.  Anyway, that contribution has been found not to
influence the collapse time appreciably, so it has been neglected in
the calculations.

From every set of 10 realizations, nearly 10000 points have been
randomly extracted to analyze the statistics of collapse times.
Figs. 2a-j and 3a-j show the scattergrams of the five collapse
estimates, for $n=-2$ and $n=1$.  Points which are predicted not to
collapse have been assigned a small negative collapse time,
-0.1. Then, the points which are predicted to collapse according to
one prediction, but not according to the other, are recognizable in
the scattergrams as horizontal or vertical rows of points. These
points will be called {\it discordant} in the following. Mean values
and dispersions around the bisector of non-discordant points are
superimposed on the scattergram.  Finally, Figs 2 and 3 focus on the
interesting zone $b_c<3$, which includes the points forming the
large-mass part of the MF.

Many conclusions can be drawn from Figs. 2 and 3:

\begin{enumerate}
\item As expected, SPH correlates with the other predictions for the
fastest collapsing points (in agreement with Bernardeau 1994), but it
badly overestimates the collapse time in general cases; moreover, many
points (those with $\delta_l<0$!) are incorrectly not predicted to
collapse.  Then, non-locality strongly accelerates collapse with
respect to the spherical case, in line with the conclusions of M95. As
a conclusion, spherical collapse is not suitable, even statistically,
for describing gravitational collapse.

\item The 1ST -- 2ND and the 2ND -- 3RD correlations at small collapse
times are increasingly good (though the former has a considerable
scatter); this demonstrates the convergence of the Lagrangian series
in predicting the collapse time of the fast collapsing points. The 1ST
-- 3RD correlation is similar to that of 1ST -- 2ND.

\item The discordant points in the 2ND -- 3RD scattergram are either
some initially slightly non-negative ones, for which 2ND does not find
any solution, as in the ellipsoidal case, (and then
$b_c^{(2ND)}$=$-$0.1), or voids which are incorrectly predicted to
collapse by 2ND (and then $b_c^{(3RD)}$=$-$0.1). This shows the
necessity of third order calculations for the collapse time. The same
features are recognizable in the other 2ND scattergrams. (N.B. The
very small number of discordant points, less than 10 in 10000, which
collapse according to 1ST and ELL, but not according to 3RD, are the
ones missed by the algorithm which looks for $J=0$; see the discussion
above.)

\item 3RD accelerates the collapse of the points with $b_c>1$ with
respect to all the other predictions. Given the uncertainties
connected with third-order calculations, this feature is not
considered very robust.  Moreover, 3RD predicts the collapse of more
points than 1ST. Again it is not clear at all if this feature,
regarding points completely outside of the convergence range of the
Lagrangian series, has some meaning.  Probably Lagrangian
perturbations are not a good means for determining the fraction of
collapsed mass in a smooth universe.

\item ELL shows an encouraging correlation with 2ND and an even better
one with 3RD, when $b_c$ is small. Note how 1ST, 2ND and 3RD, when
$b_c$ is small, converge to a solution which tightly correlates with
ELL. This has two implications, namely that the Lagrangian series
probably converges to the true solution and that ELL can be used as a
realistic estimate of the collapse time. Moreover, ELL tends to
underestimate the collapse time for $b_c>1$, slightly with respect to
2ND and strongly with respect to 3RD. This would mean that the
non-locality not contained in the ELL estimate speeds up the collapse.

\item The results are essentially independent of $n$; the weak
differences that can be visible in the scattergrams will be quantified
in the following.
\end{enumerate}

\begin{figure*}
\centerline{
\psfig{figure=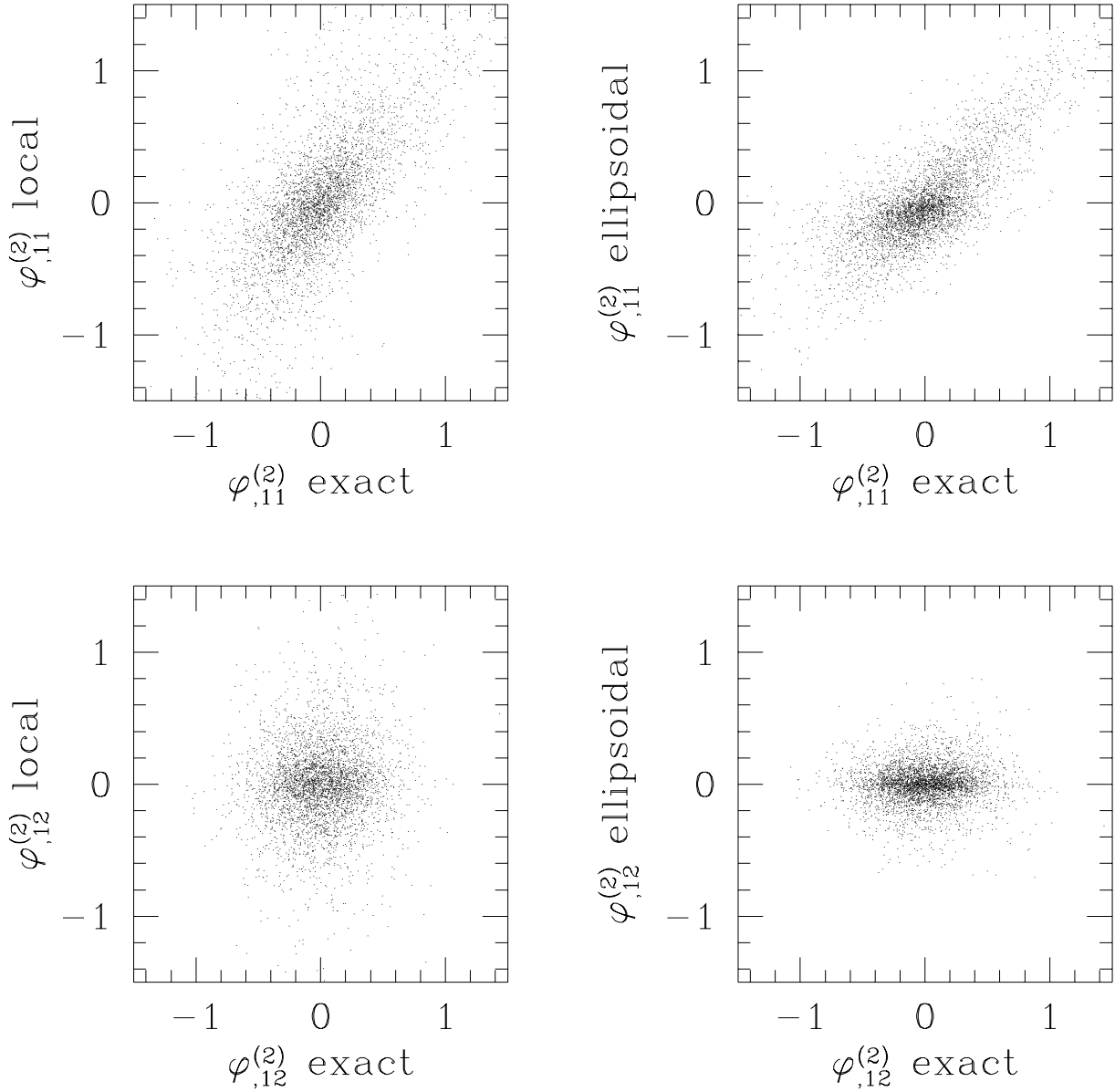,width=17cm}
}
\caption{Scattergrams of diagonal ($\phi_{,11}$) and
non-diagonal ($\phi_{,12}$) matrix elements of the 2nd-order
contributions to the deformation tensor, calculated either exactly or
by using local forms or their ellipsoidal parts.}
\end{figure*}

These results are strictly correct for an Einstein de-Sitter Universe,
in which case the scale factor can be used as a time variable,
$b_c=a_c$.  Due to the very weak $\Omega$-dependence of the time
functions, when expressed in terms of $b$, the results for $\Omega\neq
1$ cosmologies are nearly indistinguishable.  The most important
effect, in the open case, is that the growing mode saturates at the
value $b = 5/2$; a mass element with that collapse ``time'' collapses
at an infinite physical time, and larger $b_c$ imply no collapse.

\begin{figure*}
\centerline{
\psfig{figure=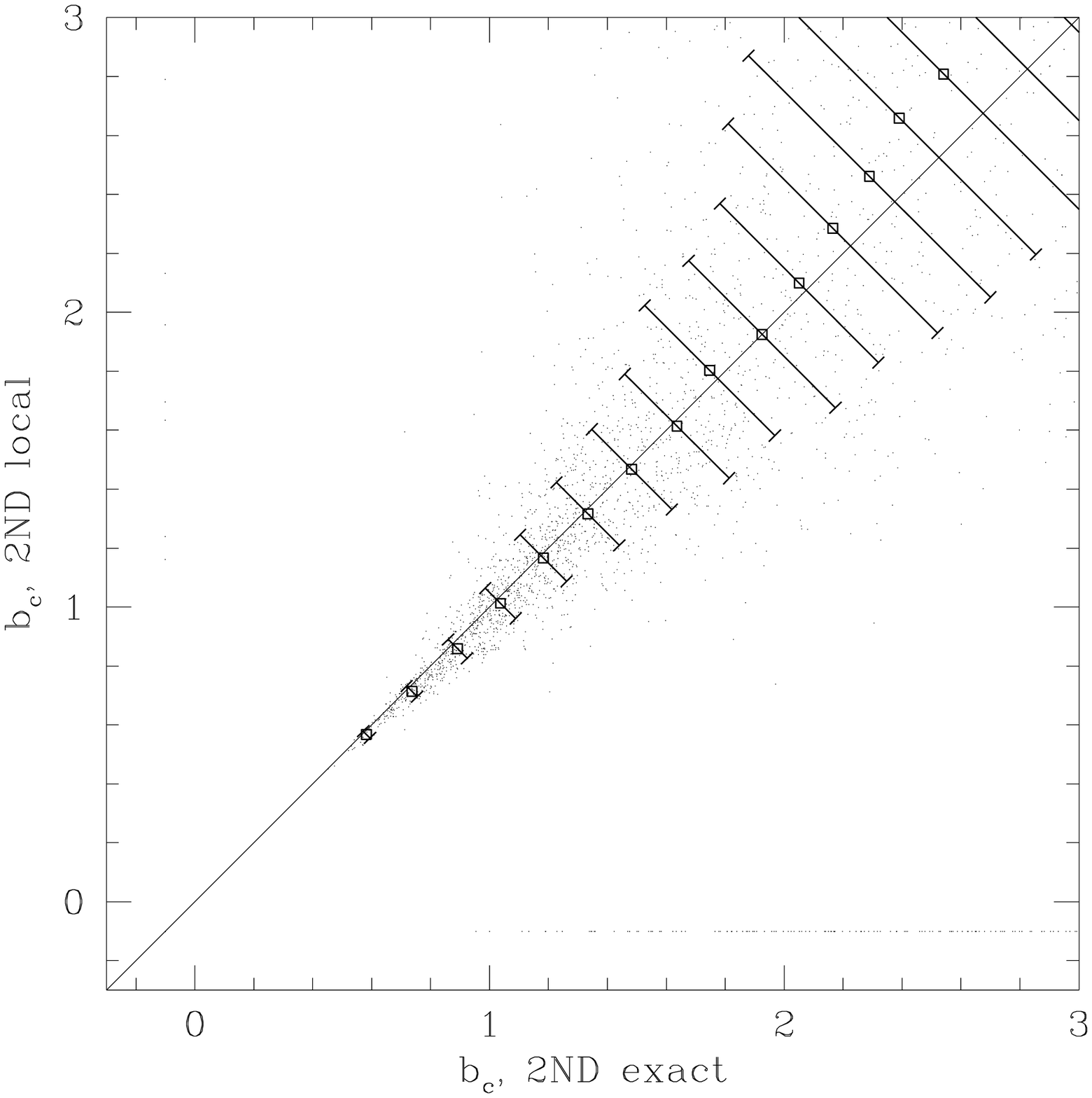,width=8cm}
\psfig{figure=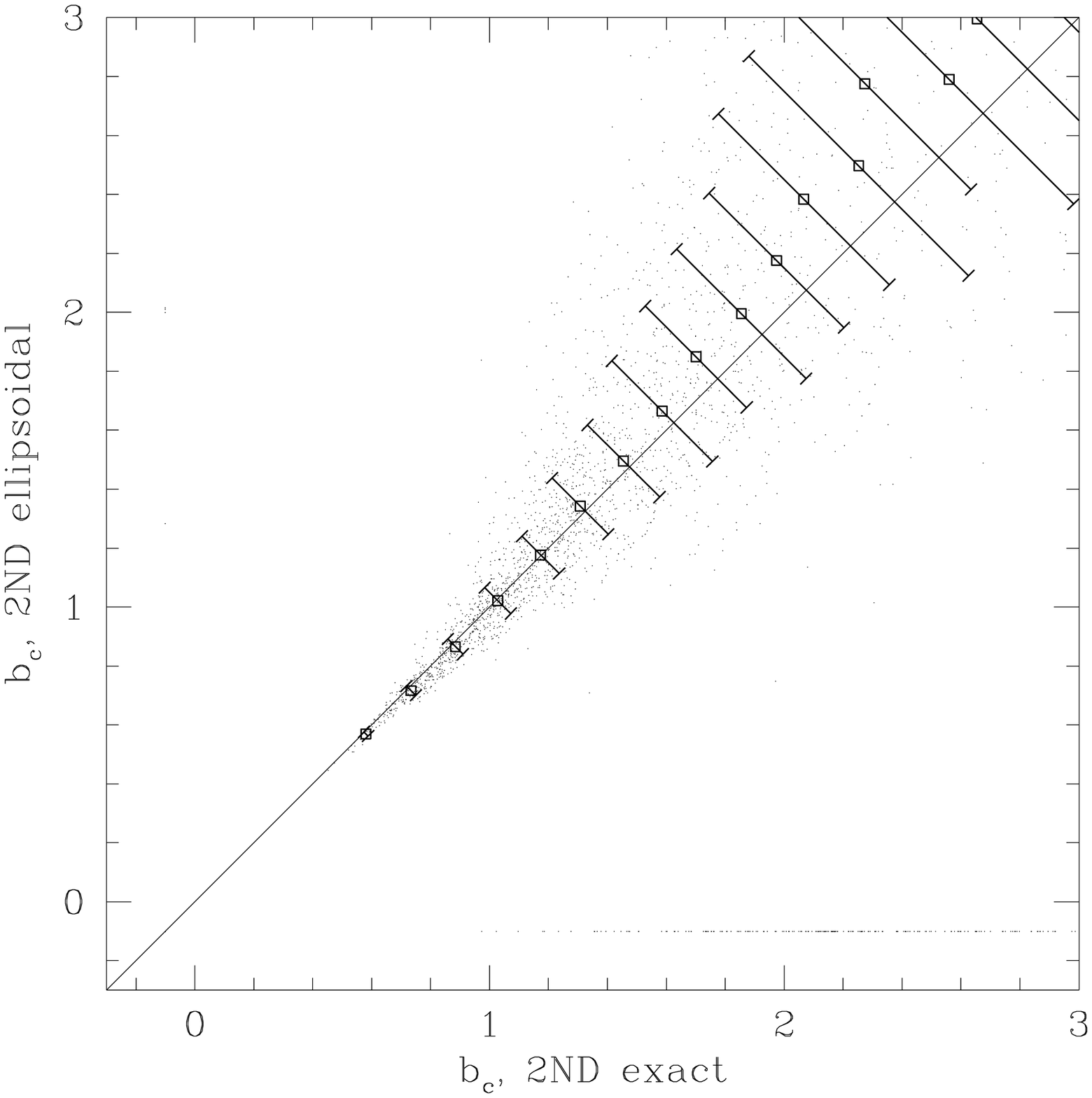,width=8cm}
}
\caption{Scattergrams of 2ND collapse time estimates
$b_c$, calculated either exactly or by using local forms or their
ellipsoidal parts.}
\end{figure*}

As noted in Section 4, ELL can be considered as a truncation of the
Lagrangian series, or, better, of its `local forms'.  It is
interesting to check whether the use of the full local forms of the
Lagrangian series improves the agreement of the approximation to the
complete calculation. Here only the second-order terms will be
considered; in this case, 2nd-order local and ellipsoidal
displacements are equal, the differences come into the deformation
tensor. The 2nd-order deformation tensor, its local and its
ellipsoidal parts have been calculated for every point of a 16$^3$
realization with power spectrum $n=0$. Fig. 4 shows the scattergrams
of one of the diagonal elements of the 2ND deformation tensor,
$S^{(2)}_{1,1}$ versus its local form, $S^{(2L)}_{1,1}$, and its
ellipsoidal part, $S^{(2E)}_{1,1}$. Fig. 4 also shows the same
pictures for one of the off-diagonal elements; other diagonal and
off-diagonal elements obviously behave identically.  The local
diagonal elements correlate with the 2ND ones just slightly better
better than the ellipsoidal ones, while no correlation of the
off-diagonal terms is visible in any case.  Fig. 5 shows the
corresponding collapse times; here the ellipsoidal collapse time is
given by equation (\ref{eq:el2}), without any correction. Again, the
ellipsoidal part reproduces the full 2ND collapse time nearly as well
as the local form.  As a conclusion, it is convenient to truncate the
local forms as in Section 4, as it leads to a great simplification of
the calculations and to a similar agreement with the full Lagrangian
terms.

\subsection{The inverse collapse time PDF}

\begin{figure*}
\centerline{
\psfig{figure=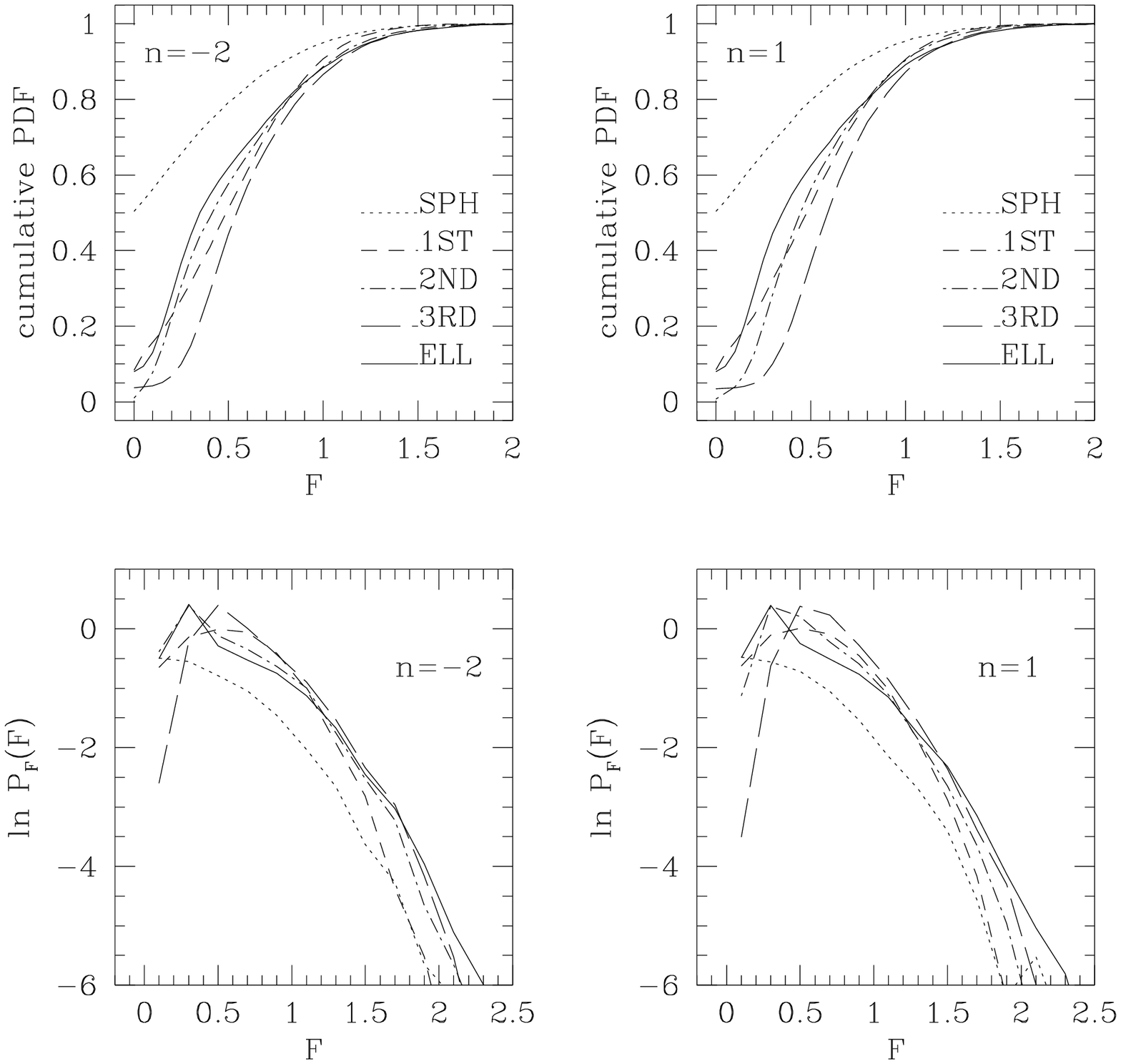,width=17cm}
}
\caption{Cumulative and differential PDF of various inverse collapse
time estimates $F$, for $n=-2$ or $n=1$.}
\end{figure*}

To quantify the PDF of the collapse times, it is convenient to
consider their inverses, as they are better behaved: the inverse
collapse times are large, but of order one, for fast collapsing
points, and become smaller and smaller for slowly collapsing ones; the
passage from collapse to non collapse is not at infinity, as
for $b_c$, but at 0. We define:

\be F(\mq) = b_c^{-1}(\mq)\; . \ee

\noindent In the SPH case, $F=\delta_l/1.69$; in the 1ST case,
$F=\lambda_1$. This definition is also convenient for
determining the MF (see paper II).

Fig. 6 shows both the cumulative and differential $F$ PDFs for $n=-2$
and 1; the cumulative curves give a binning-free picture of the PDFs,
while the differential ones, in logarithmic scale, better exhibit the
behaviour in the rare event tail.  The following things can be noted:

\begin{enumerate}
\item The SPH curve is quite different from all the others, even at
the high $F$ tail: as in M95, a systematic departure from spherical
collapse influences also the statistics of rare events.

\item In the range $F\geq1$, the 1ST, 2ND and 3RD curves show a
monotonic shift toward large $F$ values, which can be interpreted as
convergence toward a solution. This is not true for $F<1$; the bad
behaviour of 2ND in initial underdensities is the probable cause.

\item ELL and 3RD nearly coincide down to $F=1.2$, which corresponds
to $b_c=0.83$, and overall have a similar behaviour; ELL slightly makes
more mass collapse at large $F$ values, because 3RD slightly
underestimates quasi-spherical collapses.  The main differences come
out in the range where the convergence of the Lagrangian series is not
guaranteed, but both 1ST, 2ND and 3RD have larger medians than ELL; so
ELL probably underestimates the collapses around $F=0.5$ or $b_c=2$.

\item The $n$ dependence can be appreciated in Fig. 6.  As expected,
SPH, 1ST and ELL are essentially independent of $n$ (the $\lambda_i$
distribution is fixed once the mass variance is fixed), while the $n$
dependence of 2ND and 3RD is weak. Moreover, the difference between
ELL and 3RD is smaller for smaller $n$, i.e. when more large-scale
power is present, while we would expect the opposite if the difference
of 3RD from ELL were due to non-locality induced from large scales. In
the following the weak $n$-dependence of 3RD will be neglected.

\item The range $F>1$, where the Lagrangian series converges to ELL,
corresponds to more than 10 per cent of the points, i.e., of the
mass. 20 per cent of the mass is found in $F>0.8$, where ELL and 3RD
are not very different, while all the medians can be found around
$F\sim0.5$. So the convergence of the Lagrangian series and its
agreement with ELL take place for a significant quantity of mass,
covering more than the large-mass tail of the MF, while the
appreciable differences between ELL and 3RD affect the power-law part
of the MF, which is plagued by a number of other problems (see paper
II).
\end{enumerate}

\begin{figure*}
\centerline{
\psfig{figure=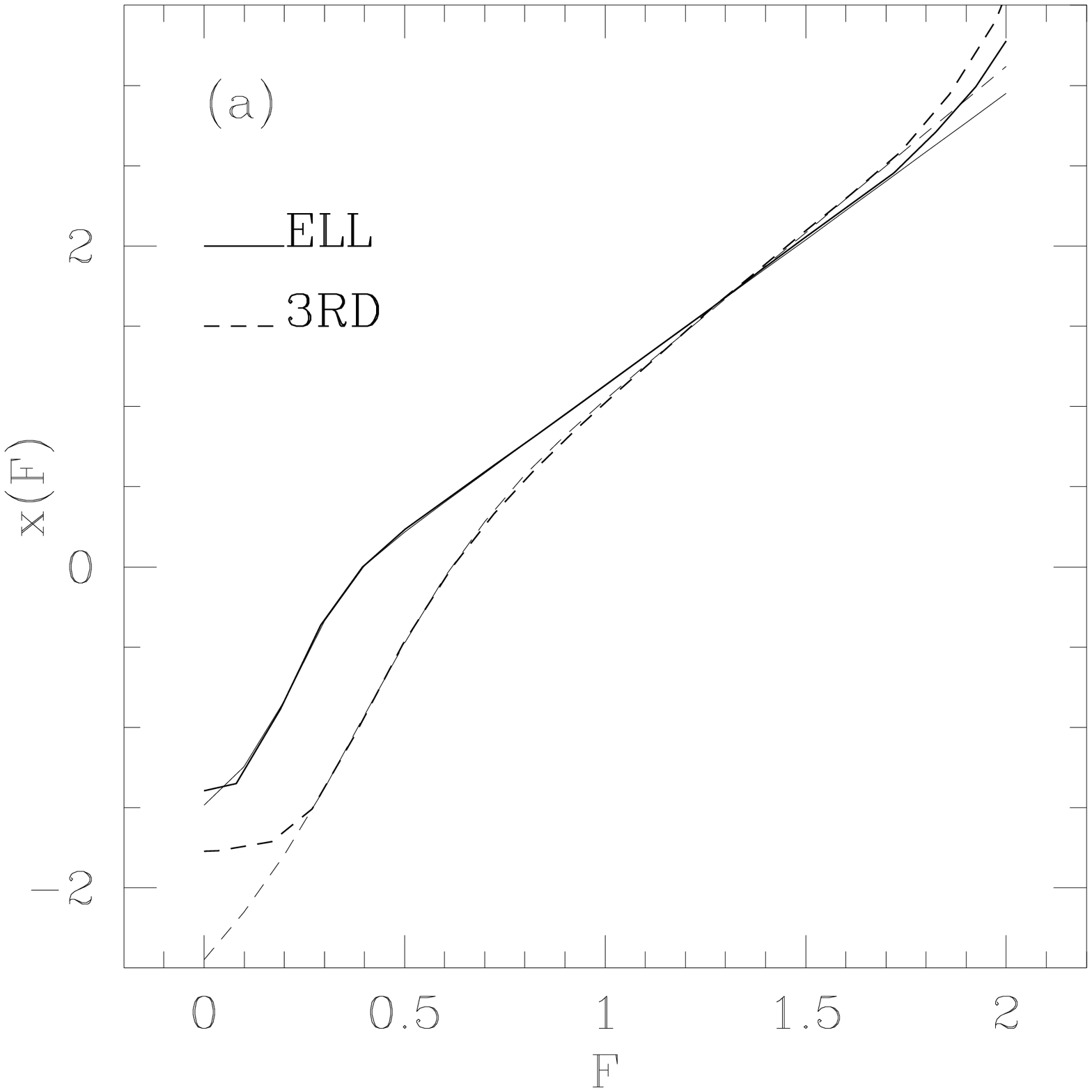,width=8cm}
\psfig{figure=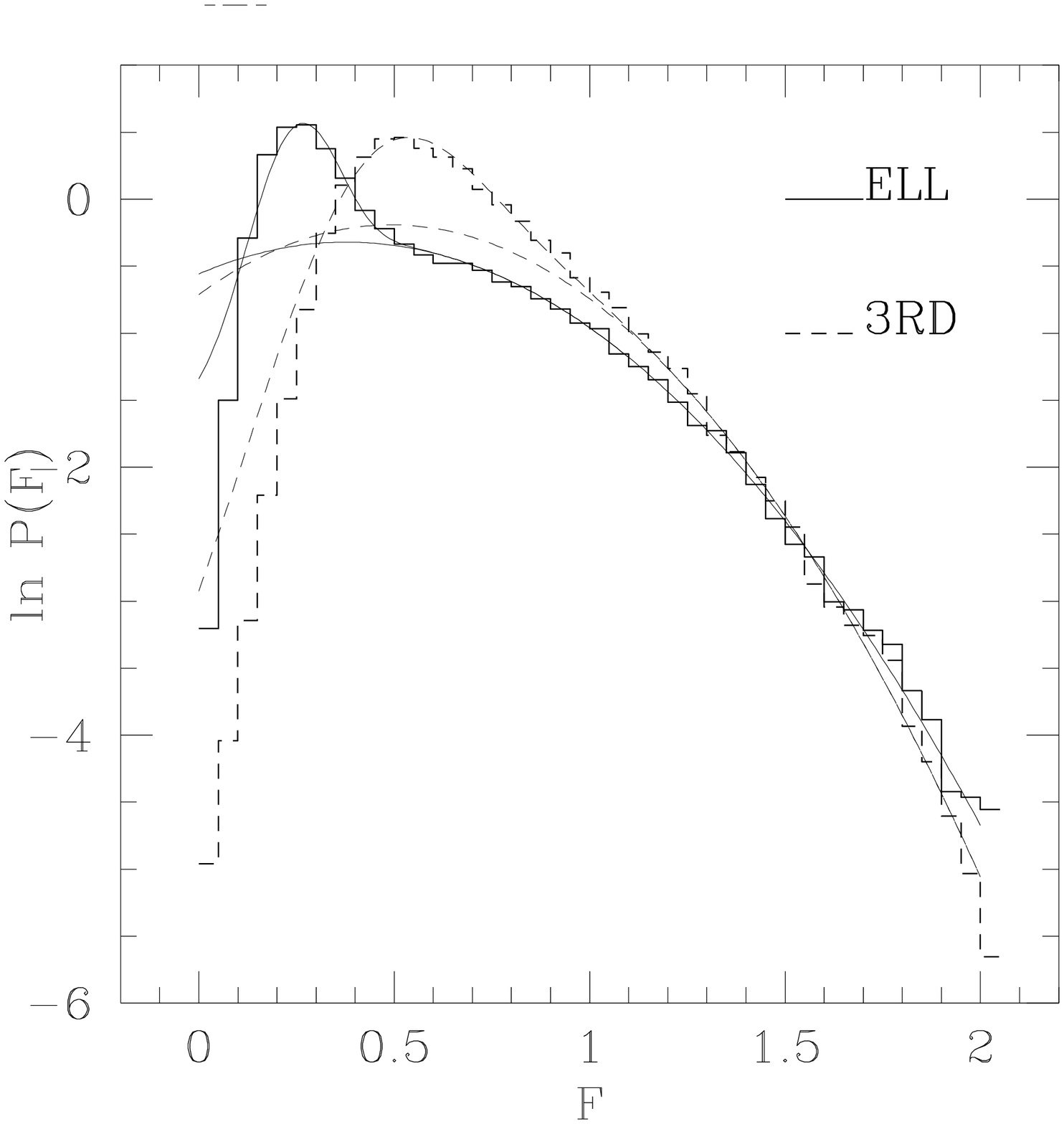,width=8cm}
}
\caption{(a): Heavy lines: transformations of variable
$x(F)$ which make the ELL or 3RD $F$ PDFs Gaussian. Light lines:
analytical approximations. (b): Histograms of 3RD and ELL $F$ PDFs,
with their Gaussian and complete analytical approximations.}
\end{figure*}

To calculate the MF, an analytical expression for the $F$ PDF is
needed.  To find it, it is useful to look for a coordinate
transformation which maps the given PDF to a Gaussian one, with zero
mean and unit variance. Consider then the quantity $F$ with its PDF
$P_F(F)$, and a normal Gaussian $P_G(x)=\exp(-x^2/2)/\sqrt{2\pi}$. If
the function $P_F$ is well-behaved enough, a transformation $x(F)$
will exist such that:

\be P_F(F)dF = P_G(x(F))dx(F) = \frac{1}{\sqrt{2\pi}}
{\rm e}^{-x(F)^2/2} dx(F)\; . \ee

\noindent The function $x(F)$ can be found as a solution of the
ordinary differential equation:

\be \frac{dx(F)}{dF} = \sqrt{2\pi} P_F(F) \exp\left(\frac{x(F)^2}{2}
\right)\; . \label{eq:tran} \ee

\noindent This equation can be easily solved with a computer. The initial
condition of the equation is best set by integrating from the median
of the $P_F$ distribution, setting $x(F)=0$ at that point.

Equation (\ref{eq:tran}) has been solved for the ELL and 3RD
distributions.  The 3RD distribution has been mediated over the four
spectral indexes considered, so as to neglect the $n$ dependence. The
results are shown in Fig. 7a: in both cases the transformations for
$F>1$ are accurately linear, i.e. the large-$F$ parts of the $F$
distributions are Gaussians.(Note that the weak rise of the $x(F)$
curve in the large-$F$ end is an artifact). The $x(F)$ transformations
are accurately fit by the following expressions:

\bea x(F)_{ELL} = -0.69 + 1.82F - 0.4 ({\rm erf}(-7.5 F + 1.75) + 1)\\
     x(F)_{3RD} = -1.02 + 2.07F - 0.75({\rm erf}( -3  F + 1.18) + 1) \; ;
\nonumber \eea

\noindent these analytical fits do not reproduce well the $x(F)$
curves for $F<0.2$; on the other hand, that part of the PDFs is very
uncertain. The first two terms represent the linear fits, valid for
large $F$ values.  Fig. 7b shows the two ELL and 3RD PDFs, together
with their Gaussian (coming from the linear transformations) and
non-Gaussian (coming from the complete transformation) fits.  Both
non-Gaussian distributions, compared with the Gaussian ones, show a
peak at small $F$; the two peaks have similar heights but slightly
different positions. These correspond to the falling tail in the
transformation curves $x(F)$.

%%%%%%%%%%%%%%%%%%%%%%%%%%%%%%%%  6  %%%%%%%%%%%%%%%%%%%%%%%%%%%%%%%%%%%%%

\section{Summary and Conclusions}

In this paper, the dynamical part of a new theory of the mass function
of cosmic structures has been presented. An accompanying paper, paper
II, develops the statistical machinery necessary to calculate the MF.
The key point of all the dynamical analysis is the use of a
fluidodynamical Lagrangian framework to predict the fate of a mass
element of a smooth field. The Lagrangian perturbative scheme has been
used to approximate the real dynamics of the smoothed cosmological
fluid up to orbit crossing. The strongest hypothesis of the whole
theory is that small-scale structure does not influence the dynamics
of larger collapsing scales. The OC instant has been chosen as
suitable definition of collapse of a mass element. This definition has
been amply discussed.

To check the performances of Lagrangian perturbations in predicting
the OC instant, when all the terms of the series are of the same
order, the collapse of a homogeneous ellipsoid has been
considered. The Lagrangian series converges fast in predicting the
collapse of a homogeneous ellipsoid, except for quasi-spherical ones;
a simple correction can be applied in this case.  Third-order terms
are necessary to give good collapse estimates of initial
underdensities.  On the other hand, ellipsoidal collapse can be
considered as a truncation of the {\it whole} Lagrangian series. In
this case, as in M95, ellipsoidal collapse has to be considered an
approximation of the local collapse of a point mass, not of an
extended region of approximate ellipsoidal shape.

The collapse of scale-free Gaussian fields simulated on a 32$^3$ grid
has been considered. The Lagrangian series has been found to clearly
converge to a solution for the fast-collapsing points, about 10 per
cent of the total mass. Approximate convergence has been observed for
about 50 per cent of the mass. The homogeneous ellipsoidal model has
been found to strongly correlate with the third-order Lagrangian
collapse in the same range of validity. The ellipsoidal model can thus
be used as a fast and easy to implement approximation of collapse
dynamics.  The spherical model, instead, has been found to behave
differently from the other predictions.  The inverse collapse-time
distributions have been calculated for the third-order Lagrangian and
ellipsoidal predictions. These will be necessary for determining the
MF.

With this dynamical description of collapse, it is possible to
construct a MF, fully based on realistic dynamics. The role of the
initial density is taken by the inverse collapse time $F$. This
quantity is by no means a Gaussian process, but is a complicated
non-linear functional of a parent Gaussian process. This introduces a
number of complications in the statistical treatment of the MF, which
will be faced in paper II.  

The following points, raised by the analysis presented here, are
relevant when constructing a MF:

\begin{enumerate}
\item As the main quantity, $F$, is (the inverse of) a time, any
threshold $F_c$ in this theory simply specifies the time at which the
MF is examined; there is no free $\delta_c$ parameter.

\item The dynamical predictions are strictly punctual; in other words,
a point collapses if it is predicted to collapse (at a given scale),
not if a neighboring collapsing point is able to involve it in its
collapse. 

\item Smoothing is necessary because of the truncated nature of the
dynamical approximations used. Thus the shape of the filter has to be
chosen in order to optimize the performances of the dynamical
predictions; usually Gaussian filters are suggested.
\end{enumerate}

The collapse time, which is needed to determine the MF, is not the
only information one can obtain on a collapsing mass element.  In
practice, all the trajectories of the collapsing points are known.
This has the consequence that much more than a MF can be
calculated. A possible application is the determination of the angular
momentum of a collapsing region; Lagrangian perturbation theory is an
appropriate framework for estimating such a quantity, as Catelan \&
Theuns (1996a,b) have recently shown. The richness of the dynamical
information makes it possible to put further dynamical constraints on
the collapsing regions, if special classes of objects are required.

To finally decide on the validity of the MF dynamics presented here,
the predictions of this theory have to be tested against N-body
simulations. As collapsed regions are defined to coincide with
orbit-crossed regions, they have to be sought for by a suitable
algorithm; equation (\ref{eq:map}) could provide such an algorithm.
This MF is expected to carefully reproduce the N-body MF as long as
Lagrangian perturbation theory is expected to give a correct
description of collapsing dynamics, i.e. for large masses, comprising
at least 10-20 per cent of mass, and when few small-scale structures are
present, i.e. with small spectral indexes. 

\section*{Acknowledgments}

I wish to thank Alfonso Cavaliere and Sabino Matarrese for a number of
discussions and for their encouragement. I also thank Thomas Buchert
and Paolo Catelan for discussions on the Lagrangian perturbation
theory, and an anonymous referee for its suggestions which have been
of great help in improving the presentation. This work has been
partially supported by the Italian Research Council (CNR-GNA) and by
the Ministry of University and of Scientific and Technological
Research (MURST).

%%%%%%%%%%%%%%%%%%%%%%%%%%%%%%%  R  %%%%%%%%%%%%%%%%%%%%%%%%%%%%%%%%

\appendix

%%%%%%%%%%%%%%%%%%%%%%%%%%%%%%  A  %%%%%%%%%%%%%%%%%%%%%%%%%%%%%%%%%%%%%

\section{Lagrangian perturbations}

This appendix contains a number of technical points on Lagrangian
perturbation theory.

Peculiar velocity, acceleration and density contrast of a mass
element can be written in terms of the \s\ field as:

\bea 
{\bf v}(\mq,t)  & = & a(t)d\ms(\mq,t)/dt \nonumber \\
{\bf g}(\mq,t)  & = & a(t)d^2\ms(\mq,t)/dt^2+2H{\bf v}(\mq,t)\label{eq:kine}\\ 
1+\delta(\mq,t) & = & J(\mq,t)^{-1} [1+\delta(\mq,t_0)]\; ; \nonumber
\eea

\noindent here ${\bf v}(\mq,t)$ is the peculiar velocity of the
element \q, {\bf g}(\mq,t) is its peculiar acceleration, $\delta$ its
density, $d/dt$ denotes total (Lagrangian) time derivative, $a(t)$ is
the scale factor of the background cosmology, $H=a^{-1}da/dt$ is the
Hubble parameter, $t_0$ is an initial time and $J(\mq,t)$ is the
Jacobian determinant, equation (\ref{eq:detjac}).

The evolution equations for the displacement field \s\ can be written,
following Catelan (1995), as:

\be [(1+{\bf \nabla}\cdot\ms)\delta_{bd} - S_{b,d} + S^C_{b,d}] 
{\ddot S}_{a,d} = \alpha(\tau)[J-1] \label{eq:lag1} \ee

\be \varepsilon_{abc}[(1+{\bf \nabla}\cdot\ms)\delta_{bd} - S_{b,d} 
+ S^C_{b,d}]{\dot S}_{c,d} = 0\; , \label{eq:lag2} \ee

\noindent where the dot denotes the Lagrangian derivative with respect
to the time variable $\tau=t^{-1/3}=a^{-2}$, if $\Omega_0=1$, or 

\be \tau = |1-\Omega|^{-1/2} \label{eq:tau} \ee

\noindent otherwise. $\varepsilon_{abc}$ is the Levi-Civita
antisymmetric tensor, ${\bf \nabla}=\partial/\partial\mq$, $\ms^C$ is
the cofactor matrix of \s\ and the function $\alpha(\tau)
=6/(\tau^2+k)$ ($k=-1$, 0 or 1 for open, flat and closed models). The
first equation (\ref{eq:lag1}) is an evolution equation for \s ,
i.e. the equation of motion of a mass element, while the second
equation (\ref{eq:lag2}) is the irrotationality condition for the
peculiar velocity ${\bf v}$ {\it in Eulerian space}. The
irrotationality condition restricts the set of solutions to the
irrotational ones, which is reasonable in our cosmological context as
any rotational mode is severely damped in the early linear evolution;
see Buchert (1992) for a detailed discussion of this point. Another
useful restriction on the solutions of equations (\ref{eq:lag1}) and
(\ref{eq:lag2}) is the initial parallelism of peculiar velocity and
acceleration, which is also supported by the growing modes in the
linear and quasi-linear regime (Buchert 1992; Buchert \& Ehlers 1993).

The perturbative expansion is performed as follows. We can write the
displacement field \s\ as:

\be \ms(\mq,t) = \varepsilon \ms^{(1)}(\mq,t) + \varepsilon^2 
\ms^{(2)}(\mq,t) + \varepsilon^3 \ms^{(3)}(\mq,t) + {\cal O}(\varepsilon^4)
\; ,\ee

\noindent where $\varepsilon$ is a small parameter. Putting this
expression into the evolution equations (\ref{eq:lag1},\ref{eq:lag2})
for \s, and considering terms of order $\varepsilon$, $\varepsilon^2$
and $\varepsilon^3$ separately, one finds equations for the various
$\ms^{(n)}$ terms. It turns out that, at any order (as recently
demonstrated by Ehlers \& Buchert 1996), the solutions are separable
in time and space. At first and second orders, the solutions are
irrotational in Lagrangian space, i.e. the matrices $S^{(1)}_{a,b}$
and $S^{(2)}_{a,b}$ are symmetric. At the third order, the equation
for $\ms^{(3)}$ is not separable as it stands, but it is possible to
divide the $\ms^{(3)}$ term into three different modes, all separable
in space and time:

\be \ms^{(3)}(\mq,t) = \ms^{(3a)}(\mq,t) + \ms^{(3b)}(\mq,t)
+ \ms^{(3c)}(\mq,t)\; . \ee

\noindent The third term, not reported by Bouchet et al. (1995) who
consider only those terms which contribute to $S_{a,a}$, is a purely
rotational mode in Lagrangian space ($S^{(3c)}_{a,b}$ is
antisymmetric), and its existence is necessary to guarantee
irrotationality in Eulerian space. This fact can be understood in this
way: the Lagrangian to Eulerian transformation is in general a
non-Galileian one, so the rotational mode can be seen as an effect of
inertial forces (see the discussions in Buchert 1994 and Catelan
1995).

Perturbative terms are listed in the following. The equations for the
time functions $b_n$ contain both growing and decaying modes, which
have to be consistently considered in the calculations; however, for
present purposes only the growing modes for each $b_n$ are needed.
The time functions are accurately described by the following
expressions (exact for the first order):

\bea
b_1 & = & -b(t)\nonumber\\ 
b_2 & = & -\frac{3}{14} b_1^2 \Omega^{-a} \nonumber \\
b_{3a} & = & \frac{1}{9} b_1^3 \Omega^{-b} \label{eq:tpert}\\
b_{3b} & = & -\frac{5}{42} b_1^3 \Omega^{-c} \nonumber \\
b_{3c} & = & \frac{1}{14} b_1^3 \Omega^{-d}\; , \nonumber 
\eea

\noindent where $b(t)$ is the linear growing mode. In an Einstein-de
Sitter background it is simply:

\be b(t) = a(t)\; . \ee

\noindent In an open Universe, it is convenient to express the growing
mode in terms of the time variable $\tau$, equation (\ref{eq:tau}):

\be b(\tau) =  \frac{5}{2}\left( 1+3(\tau^2-1) \left( 1+\frac{\tau}{2} \ln 
\left( \frac{\tau-1}{\tau+1} \right)\right)\right)\; .\ee

\noindent The growing mode has been normalized so as to give $b(t)
\simeq a(t)$ at early times.  In a flat Universe with cosmological
constant, it is useful to use $h={\rm coth} (3H_0t\sqrt{1-
\Omega_0}/2)$ as time variable:

\be b(h) = h \int_h^\infty (x^2(x^2-1)^{1/3})^{-1}dx\; . \ee

The coefficients $a$, $b$ and $c$ in equation (\ref{eq:tpert}) have
been calculated in Bouchet et al. (1992) and Bouchet et al. (1995),
and are $a$=2/63, $b$=4/77 and $c$=2/35 in the non-flat cases,
$a$=1/143, $b$=4/275 and $c$=269/17875 in the flat $\Lambda\neq 0$
cases.  The above-cited authors have not estimated the
$\Omega$-dependence of the 3c term, as they do not take that term into
account. However, as clarified in the text, the 3c term can be safely
neglected.  It can be appreciated that the $b_2/b^2$ and $b_3/b^3$
terms weakly depend on $\Omega$ when $\Omega\sim 1$; then, in a
Universe with $\Omega_0\geq 0.2$, as our Universe appears to be, the
$\Omega$ dependence in equations (\ref{eq:tpert}) can be safely
neglected.

The spatial equations for the $\ms^{(n)}(\mq)$ terms are Poisson
equations. It is convenient to express $\ms^{(1)}$, $\ms^{(2)}$,
$\ms^{(3a)}$ and $\ms^{(3b)}$ in terms of scalar potentials, and
$\ms^{(3c)}$ in terms of a vector potential:

\bea
\ms^{(n)}  & = & {\bf \nabla} \varphi^{(n)},\ \ n=1,2,3a,3b  \label{eq:ppot} \\
\ms^{(3c)} & = & {\bf \nabla} \times \bvarphi^{(3c)} \; .
\nonumber\eea

\noindent Defining the principal and mixed invariants of one or two
tensors as follows:

\bea
\mu_1(A_{ab}) & = & {\rm tr}(A_{ab})=A_{aa}\nonumber \\
\mu_2(A_{ab},B_{ab}) & = & \frac{1}{2}(A_{aa}B_{bb}-A_{ab}B_{ab}) \\
\mu_2(A_{ab}) & = & \mu_2(A_{ab},A_{ab}) \nonumber\\
\mu_3(A_{ab}) & = & \det(A_{ab}) \nonumber 
\eea

\noindent (note that $\mu_1(\varphi_{,ab})\equiv\nabla^2\varphi$), it
is possible to write the following equations for the potentials:

\bea
\varphi^{(1)} & = & \varphi \nonumber\\
\nabla^2\varphi^{(2)} & = & 2\mu_2(\varphi^{(1)}_{,ab}) \nonumber\\
\nabla^2\varphi^{(3a)} & = & 3\mu_3(\varphi^{(1)}_{,ab})\label{eq:spert}  \\
\nabla^2\varphi^{(3b)} & = & 2\mu_2(\varphi^{(1)}_{,ab},\varphi^{(2)}_{,ab})
\nonumber \\
\nabla^2\varphi^{(3c)}_a & = & \varepsilon_{abc}\varphi^{(1)}_{,cd}
\varphi^{(2)}_{,db}\; . \nonumber\eea

\noindent The first equality is a consequence of initial conditions.
Catelan (1995) gives expressions for the Fourier transforms of the
solutions of all these equations, which are useful for calculating,
e.g., mean values of the perturbing terms.

%%%%%%%%%%%%%%%%%%%%%%%%%%%%%%%  B  %%%%%%%%%%%%%%%%%%%%%%%%%%%%%%%%%%%%%

\section{Ellipsoidal collapse}

Let $\varphi({\bf q})$ be a quadratic potential in its principal
reference frame:

\be \varphi({\bf q}) = \frac{1}{2}(\lambda_1 q_1^2 + \lambda_2 q_2^2 +
\lambda_3 q_3^2)\; . \ee

It is easy to calculate all the perturbative terms in this case,
especially if the local forms are used (see Buchert \& Ehlers 1993,
Buchert 1994 and Catelan 1995 for further details):

\bea
\varphi^{(2L)}_{,a}&=&\varphi_{,a}\varphi_{,bb}-\varphi_{,ab}\varphi_{,b}
\nonumber\\
\varphi^{(3aL)}_{,a}&=&\varphi_{,ab}^C\varphi_{,b}\label{eq:loc2} \\
\varphi^{(3bL)}_{,a}&=&1/2(\varphi_{,a}\varphi^{(2)}_{,bb}-\varphi_{,b}
                     \varphi^{(2)}_{,ab}+\varphi^{(2)}_{,a}\varphi_{,bb}-
                     \varphi^{(2)}_{,b}\varphi_{,ab})\nonumber\\
\varphi^{(3cL)}_{a}&=&1/2(\varphi_{,b}\varphi^{(2)}_{,ab}-
                        \varphi^{(2)}_{,b}\varphi_{,ab})\nonumber\; ;
\eea

\noindent $\varphi_{,ab}^C$ is the cofactor matrix of
$\varphi_{,ab}$. These local parts are exact solutions in our case, as
they are irrotational; their expressions can be considerably
simplified, as all the derivatives beyond the second vanish. The
outcoming contributions to the deformation tensor have been given in
equation (\ref{eq:elp}). A technical remark on the 3b contribution:
Buchert (1994) divides this contribution into two parts, weighted by
two coefficients whose sum is equal to one (see his equations
27). These two coefficients could be varied in order to make the whole
contribution irrotational. On the other hand, Catelan gives an
expression analogous to the one given here (see his equation 44); this
would correspond to a choice of 1/2 for the two coefficients of
Buchert. As a matter of fact, the two parts identified by Buchert are
irrotational by themselves in the ellipsoidal case, and have the same
divergence, but are nonetheless different. Using only one or another,
which would correspond to setting one coefficient to 1 and the other
to 0, makes the Lagrangian series no to converge any more to the
numerical solution. So, the choice of 1/2 for both coefficients seems
the right one for ellipsoidal collapse.

All the contributions to the deformation tensor are diagonal in the 
same frame; the 3c contribution obviously vanishes. The diagonal (1,1)
components are:

\bea
\varphi_{,11} &=& \lambda_1\nonumber\\
\varphi^{(2)}_{,11} &=& \lambda_1(\lambda_2+\lambda_3)\\
\varphi^{(3a)}_{,11} &=& \lambda_1\lambda_2\lambda_3\nonumber\\
\varphi^{(3b)}_{,11} &=& \lambda_1\lambda_2\lambda_3+\lambda_1\delta_l
                         (\lambda_2+\lambda_3)/2\; ,\nonumber
\eea

\noindent where $\delta_l=\lambda_1+\lambda_2+\lambda_3$. The Jacobian
determinant vanishes when one of its eigenvalues vanishes. Then, if
the $\Omega$ dependence of the time functions is neglected, it is
possible to write the $J=0$ equation as a third-order algebraic
equation:

\bea \lefteqn{1-\lambda_i b_c - \frac{3}{14} \lambda_i (\delta_l-\lambda_i)
b_c^2 - \left( \frac{\mu_3}{126} + \frac{5}{84}\lambda_i
\delta_l(\delta_l-\lambda_i)\right)b_c^3}\nonumber\\ 
&&= 0\; , \eea

\noindent where $b$ is the linear growing mode, and
$\mu_3=\lambda_1\lambda_2\lambda_3$. Note how higher-order
coefficients become increasingly smaller.

In the spherical case, the equation reduces to:

\be 1-\frac{1}{3}(\delta_lb_c)-\frac{1}{21}(\delta_lb_c)^2-
\frac{23}{1701}(\delta_lb_c)^3=0\; , \ee

\noindent which, if truncated at first, second, or third-order,
gives the solutions:

\bea b_c^{(1)} &=& 3/\delta_l\nonumber \\ b_c^{(2)} &=& 2.27/\delta_l\\
     b_c^{(3)} &=& 2.05/\delta_l\; . \nonumber\eea

The complete equation at first-order gives the well known Zel'dovich
approximation:

\be b_c=1/\lambda_i\; .\ee

\noindent It is apparent that the 1-axis, corresponding to the largest
$\lambda$ eigenvalue, is the first to collapse.  The second-order
solution is:

\be b_c^{(2)} = \frac{7\lambda_1 - \sqrt{7 \lambda_1(\lambda_1+6\delta_l)}}
{3\lambda_1(\lambda_1-\delta_l)}\; . \label{eq:secordel} \ee

\noindent This solution is limited to $\delta_l\geq -\lambda_1/6$,
i.e. to overdensities or weak underdensities.  The other solution of
the equation (with a plus in front of the square root) is either
negative or larger than the chosen one, except when $\delta_l
<\lambda_i<0$. In this case, which includes spherical voids, the
second solution incorrectly predicts collapse. The bad behaviour of
second-order perturbations in predicting the dynamics of voids had
already been noted by Sahni \& Shandarin (1995). Finally, it is
possible to verify, by differentiating equation (\ref{eq:secordel})
with respect to the $\lambda_i$ parameter, that the 1-axis is the
first to collapse.

To obtain the third-order solution, let $y=\delta_l/
\lambda_i$, and $D=\mu_3\lambda_i^3/126 +5y(y-1)/84$; then let
$$Q=(3(y-1)^2-196D)/588D^2$$ 
$$R=(2(y-1)^3-196(y-1)D-2744D^2)/5488D^3$$
When $R^2-Q^3>0$, which is valid for spherical and quasi spherical
perturbations, the solution is:

\bea \lefteqn{b_c = -\frac{{\rm sgn}(R)}{\delta_l} \left( \left(\sqrt{R^2-Q^3}
+|R| \right)^{1/3}+ \right. }\\
&& \left. Q \left(\sqrt{R^2-Q^3}+|R|\right)^{-1/3} \right)-\frac{(y-1)}{14D}
\; . \nonumber\eea

\noindent In this case, it is possible to show analytically that the
1-axis is the first to collapse.  Otherwise, the solution has to be
chosen as the smallest non-negative one between:

\bea b_{c1}&=&-2\sqrt{Q}\cos(\theta/3)-(y-1)/14D \nonumber\\
b_{c2}&=&-2\sqrt{Q}\cos((\theta+2\pi)/3)-(y-1)/14D \\
b_{c3}&=&-2\sqrt{Q}\cos((\theta+4\pi/3)-(y-1)/14D \; ,\nonumber\eea

\noindent where $\theta={\rm arccos}(R/\sqrt{Q^3})$. It has been
checked with the computer that the 1-axis is the first to collapse.

The evolution equations of the homogeneous ellipsoid, which have been
numerically integrated, are analogous to the ones in M95: let
$r_i(t)=a_i(t)q_i$ be the physical $i$-th coordinate of the outer surface
of the ellipsoid, in the principal-axes system; $q_i$ is the initial
-- Lagrangian -- position and $a_i$ is the expansion factor for the
$i$-th axis. The universal scale factor is called $a$, without pedix. 
Then the evolution equations for the $a_i$ factors are:

\bea \lefteqn{\frac{d^2a_i}{da^2} - (2a(1+(\Omega_0^{-1}-1)a))^{-1}
\frac{da_i}{da}}+ \\&&(2a^2(1+(\Omega_0^{-1}-1)a))^{-1}a_i \left[ \frac{1}{3} 
+\frac{\delta}{3} + \frac{b'_i}{2} \delta + \lambda'_{vi} \right] \nonumber\eea

\noindent in the open case and 

\bea \lefteqn{\frac{d^2a_i}{da^2} - \frac{1-2(\Omega_0^{-1}-1)a^3}
{2a(1+(\Omega_0^{-1}-1)a)}\frac{da_i}{da} +}\\&&
(2a^2(1+(\Omega_0^{-1}-1)a))^{-1}a_i \left[ \frac{1}{3} +
\frac{\delta}{3} + \frac{b'_i}{2} \delta + \lambda'_{vi} \right] \nonumber\eea

\noindent in the flat case with cosmological constant. Here ($a_0$ is
the initial scale factor):

\be \delta = \frac{a^3}{a_1a_2a_3}-1 \ee

\be b'_i=\frac{2}{3} [a_i a_j a_k R_D(a_i^2,a_j^2,a_k^2)-1]\;\;\;\;\;
i\neq j\neq k \ee

\be \lambda'_{vi} = -\frac{a}{a_0} \left(\frac{\delta}{3} - 
a_0 \lambda_i \right) \ee

\be R_D(x,y,z) = \frac{3}{2}\int_0^\infty \frac{d\tau}{(\tau+x)^{1/2}
(\tau+y)^{1/2}(\tau+z)^{3/2}}\; . \ee

\noindent This Carlson's elliptical integral has been calculated by
means of the routine given by Press \& Teukolsky (1990).

The initial conditions are: 

\be a_i(a_0)=a_0(1-a_0\lambda_i) \ee
\be \frac{d a_i}{da}(a_0) = \frac{1}{a_0} (a_i(a_0)-a_0^2 \lambda_i)\; . \ee

\bsp

\label{lastpage}

\end{document}